\documentstyle[aps,prd,preprint,floats,epsfig,tighten]{revtex}

\begin{document}
\vspace*{-2cm}
\renewcommand{\thefootnote}{\fnsymbol{footnote}}
\begin{flushright}
hep-ph/9906324\\
SLAC--PUB--8019\\
DTP/98/94\\
June, 1999\\
\end{flushright}
\vskip 65pt
\begin{center}
{\Large \bf The Two-Loop Scale Dependence of the  Static QCD
Potential including Quark Masses\footnote{Work supported in part by the
Department of Energy, contract
DE--AC03--76SF00515, EU Fourth Framework Programmme 
`Training and Mobility of Researchers', 
and the Swedish Natural Science Research Council,
contract F--PD 11264--301.}}\\
\vspace{1.2cm}
{\bf
Stanley  J.~Brodsky${}^1$\footnote{sjbth@slac.stanford.edu},
Michael Melles${}^2$\footnote{Michael.Melles@durham.ac.uk} and
Johan Rathsman${}^{1}$\footnote{rathsman@slac.stanford.edu}
}\\
\vspace{10pt}
{\sf 1) Stanford Linear Accelerator Center,
Stanford University, Stanford, California 94309\\

2) Department of Physics, University of Durham,
Durham DH1 3LE, U.K.}
\end{center}

\begin{abstract}
The interaction potential $V(Q^2)$ between static test charges can be used to
define an effective charge $\alpha_V(Q^2)$ and a physically-based
renormalization scheme for quantum chromodynamics and other gauge theories.  In
this paper we use recent results for the finite-mass fermionic corrections to
the heavy-quark potential at two-loops to derive the next-to-leading order term
for the Gell Mann-Low function of the $V$-scheme. The resulting effective
number of flavors $N_F(Q^2/m^2)$ in the $\alpha_V$ scheme is determined as a
gauge-independent and analytic function of the ratio of the momentum transfer
to the quark pole mass. The results give automatic decoupling of heavy quarks
and are independent of the renormalization procedure.  
Commensurate scale  relations
then provide the next-to-leading order connection between all perturbatively
calculable  observables to the analytic and gauge-invariant
$\alpha_V$ scheme without any scale ambiguity and a well defined number of
active flavors. The inclusion of the finite quark mass effects in the running
of the coupling is compared with the standard treatment of finite
quark mass effects in the $\overline{\mbox{MS}}$ scheme.
\end{abstract}
\vfill

\setcounter{footnote}{0}
\renewcommand{\thefootnote}{\arabic{footnote}}

\vfill
\clearpage
\setcounter{page}{1}
\pagestyle{plain}

\section{Introduction}

Is there a preferred effective charge which should be used to characterize the
coupling strength in QCD? In principle, any perturbatively calculable observable
can be used to define an  effective charge. In quantum electrodynamics, the
Dyson running coupling
$\alpha_{\mbox{\tiny QED}}(Q^2)$, defined from
the linearized 
potential between two infinitely-heavy test charges, has been
used as the traditional coupling.
The corresponding  definition of the non-abelian QCD coupling
is customarily given by identifying the ground state energy of the vacuum
expectation value of the Wilson loop as
the potential $V$ between a static quark-antiquark pair in a
color singlet state \cite{Susskind}:
\begin{equation}
V( r, m^2) = - \lim_{t \rightarrow \infty} \frac{1}{it}
\log \langle 0| Tr \; \left\{P \; \exp \left(
\oint dx_\mu A^\mu_a T^a \right) \right\} |0 \rangle
\label{eq:Vdef}
\end{equation}
where $r$ denotes the relative distance between the heavy quarks, $m$ is 
the mass of ``light'' quarks contributing through loop effects, and $T^a$ are
the generators
of the gauge group. It is then convenient to define the effective charge
$\alpha_V( Q^2,m^2)$ as
\begin{equation}
V(Q^2,m^2) \equiv - 4 \pi C_F\frac{ \alpha_V ( Q^2,m^2)}{{Q}^
2}
\label{eq:aVdef}
\end{equation}
in momentum space. The factor $C_F$ is the value of the Casimir operator $T^a
T^a$ of the external sources (which are in the fundamental representation)
 and factors out to all orders in perturbation theory, and $Q^2=-q^2$ 
 is the spacelike  momentum transfer.

The effective charge $\alpha_V(Q)$ provides a physically-based alternative to
the usual modified  minimal
subtraction ($\overline{\mbox{MS}}$) scheme. As in the corresponding case of
Abelian QED, the scale $Q$ of the coupling $\alpha_V(Q)$ is given by the
exchanged  momentum. There is thus no ambiguity in the interpretation of
the scale.   All virtual corrections due to massive fermion pairs are
incorporated in  $\alpha_V$ through loop diagrams which
depend on the physical mass thresholds.  When continued to time-like momenta,
the coupling has the correct analytic dependence dictated by the production
thresholds in the crossed channel. Since  $\alpha_V$ incorporates  quark
mass effects exactly, it avoids the problem of explicitly computing and
resumming quark mass corrections which are related to the running of the 
coupling. Thus the effective number of flavors $N_F(Q/m)$ is an analytic 
function of the scale $Q$ and the quark masses $m$. The effects
of finite quark mass corrections on the running of
the strong coupling were first considered 
by De R{\' u}jula and Georgi~\cite{derujula}
 within the momentum subtraction schemes (MOM)
(see also \cite{Georgi_Politzer,Ross,shirkov,chyla}). 
The two-loop calculation was first done by Yoshino and Hagiwara~\cite{yh}
in the MOM-scheme using Landau gauge and also recently by
Jegerlehner and Tarasov~\cite{jt} using background field gauge.

One important advantage of the physical charge approach is
its inherent gauge invariance to all orders in perturbation theory. This
feature is not manifest in massive $\beta$-functions defined in non-physical
schemes such as the MOM schemes. A second, more practical,
advantage is the automatic decoupling of heavy quarks according to the
Appelquist-Carazzone theorem \cite{ac}.

By employing the commensurate scale relations~\cite{csr} 
other physical observables
can be expressed in terms of  the analytic 
coupling $\alpha_V$ without scale or scheme ambiguity. 
The quark mass threshold effects 
in the running of the coupling are taken into account by
 utilizing the mass dependence of the physical
$\alpha_V$ scheme. In effect, quark thresholds are treated analytically to all
orders in $m^2/Q^2$; {\it i.e.},  the evolution of the physical $\alpha_V$ 
coupling in the intermediate regions reflects the actual mass dependence of a
physical effective charge and the analytic properties of particle production.
  Furthermore, the definiteness of the
dependence in the quark masses automatically constrains the 
scale $Q$ in the argument of the coupling.  
There is thus no scale ambiguity in perturbative expansions in $\alpha_V$.

In the conventional $\overline{\mbox{MS}}$ scheme,
the coupling is independent of the
quark masses since the quarks are treated as either massless or
infinitely heavy with respect to the running  of the coupling. Thus one
formulates different effective theories depending on  the effective number of
quarks which is governed by the scale $Q$; the  massless $\beta$-function
is used to describe the running in between the flavor thresholds.
These different theories are then matched to each other by imposing matching
conditions at the scale (normally the quark  masses)
where the effective number of flavors is changed.  
The dependence on the matching scale
 can be made  arbitrarily small by calculating the matching conditions
to high enough order. For physical observables one can then include the effects
of finite quark masses by making a higher-twist expansion in $m^2/Q^2$ and
$Q^2/m^2$ for light and heavy quarks, respectively.  These higher-twist
contributions have to be calculated for each observable separately, so
that in principle one requires an all-order resummation of the
mass corrections to the effective Lagrangian to give correct
results. 

The specification of the coupling and renormalization scheme also depends
on the definition of the quark mass.  In contrast to QED where the
on-shell mass provides a natural definition of lepton masses,  an
on-shell definition for quark masses is complicated by the confinement
property of QCD. In this paper we will use the pole mass $m$
which has the advantage of being scheme and renormalization-scale invariant.

A technical complication of massive schemes  is that one cannot easily obtain
analytic solutions of renormalization group equations to the massive $\beta$
function, and the Gell-Mann Low function is scheme-dependent even at lowest
order.

In this paper we present a two-loop analytic extension of the
$\alpha_V$-scheme based on the recent results of Ref.~\cite{melles98}.  
The mass effects are in principle treated
exactly to two-loop order and are only limited in practice by
the uncertainties from numerical integration.
The desired features of gauge invariance and decoupling are manifest in
the  form of the two-loop Gell-Mann Low function, and we give a simple
fitting-function which interpolates smoothly the exact two-loop results
obtained by using 
the adoptive Monte Carlo integrator VEGAS\cite{vegas}.  Strong
consistency checks of the results are performed by comparing the Abelian
limit to the well known QED results in the on-shell scheme. In addition,
the massless as well as the decoupling limit are reproduced exactly, and
the two-loop Gell-Mann Low function is shown to be renormalization scale 
($\mu$) independent.

As an application we show how the analytic $\alpha_V$-scheme can be used
to calculate the non-singlet hadronic width of the Z-boson, including
finite quark mass corrections from the running of the coupling and
compare with the standard treatment in the $\overline{\mbox{MS}}$ scheme
where the corresponding effects are calculated as higher twist corrections.

Recently (see Ref.~\cite{bgmr}) we proposed an alternative way
 of incorporating mass effects connected with the running of the
 coupling by making an analytic extension
 of the $\overline{\mbox{MS}}$-scheme where the coupling is an
analytic function of both the scale  and the quark masses. This analytic
extension of  the $\overline{\mbox{MS}}$-scheme is defined by connecting the
$\overline{\mbox{MS}}$ coupling to the $V$-scheme using a commensurate scale
relation based on the  (BLM) scale-setting procedure\cite{blm}. The new
modified coupling $\widetilde {\alpha}_{\overline{\mbox{\tiny MS}}}(Q)$
inherits most of the good properties of the $\alpha_V$ scheme, including its
correct analytic properties as a function of the quark masses and its
unambiguous scale fixing \cite{bgmr}.

However, the conformal coefficients in the commensurate scale
relation between the $\alpha_V$ and
$\overline{\mbox{MS}}$ schemes do not preserve one of the
 defining criterion 
of the potential expressed in the bare charge, namely the non-occurrence of
color factors corresponding to an iteration of the potential. This
is probably an effect of the breaking of conformal invariance by
the $\overline{\mbox{MS}}$ scheme. The breaking of conformal symmetry
has also been observed when dimensional regularization is used as a
factorization scheme in both exclusive~\cite{Frishman,Muller} and
inclusive~\cite{Blumlein} reactions.  Thus, it does not turn out to be
possible to extend the modified scheme
${\widetilde \alpha}_{\overline{\mbox{\tiny MS}}}$ beyond leading order
without running into an intrinsic contradiction with conformal
symmetry.   Note, however, that this difficulty does not affect using  the
$\overline{\mbox{MS}}$ scheme as an intermediate renormalization
scheme when connecting physical observables.
For completeness we 
give the results of such an extension in an appendix.

The paper is organized as follows:
In section \ref{sec:gml} we derive the second term of the Gell-Mann Low function
in the physical $V$-scheme as a renormalization-scale-independent function
of the ratio of the physical momentum transfer $Q$ and the pole mass $m$.
In section \ref{sec:nr} we
present numerical results of the effective
number of flavors and compare it with results obtained in the
gauge-dependent momentum subtraction schemes.
In addition, various consistency checks are
performed, and numerical fits are presented.
In section \ref{sec:prop} we illustrate some of the properties
of  the analytic $\alpha_V$ scheme and
demonstrate the effect of the quark mass
thresholds on the mass-dependent evolution
and compare with the massless evolution.
In section \ref{sec:appl} we compare the calculation of the
hadronic width of the $Z$-boson in the analytic $\alpha_V$ scheme 
to the conventional 
$\overline{\mbox{MS}}$ scheme with a mass-independent coupling and
explicit higher-twist corrections for mass effects.
In section \ref{sec:sum} we summarize our results and indicate future
applications.
The definition of 
the analytically extended scheme 
${\widetilde \alpha}_{\overline{\mbox{\tiny MS}}}$
beyond leading order is discussed in the  appendix.

\section{The Gell-Mann Low Function Through Two Loops}\label{sec:gml}

\begin{figure}
\center
\epsfig{file=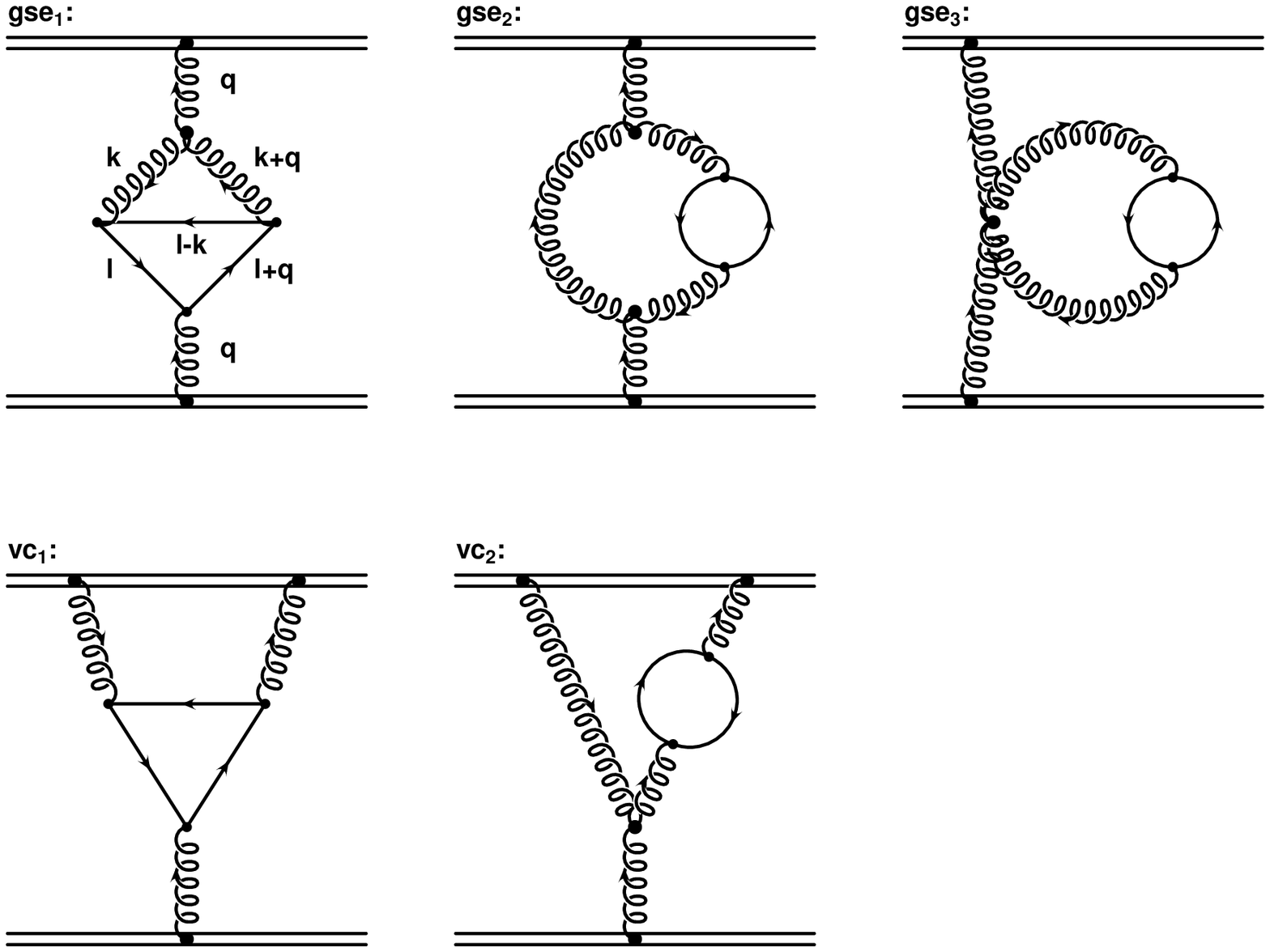,height=2.5in}
\vspace{0.5cm} \\
\epsfig{file=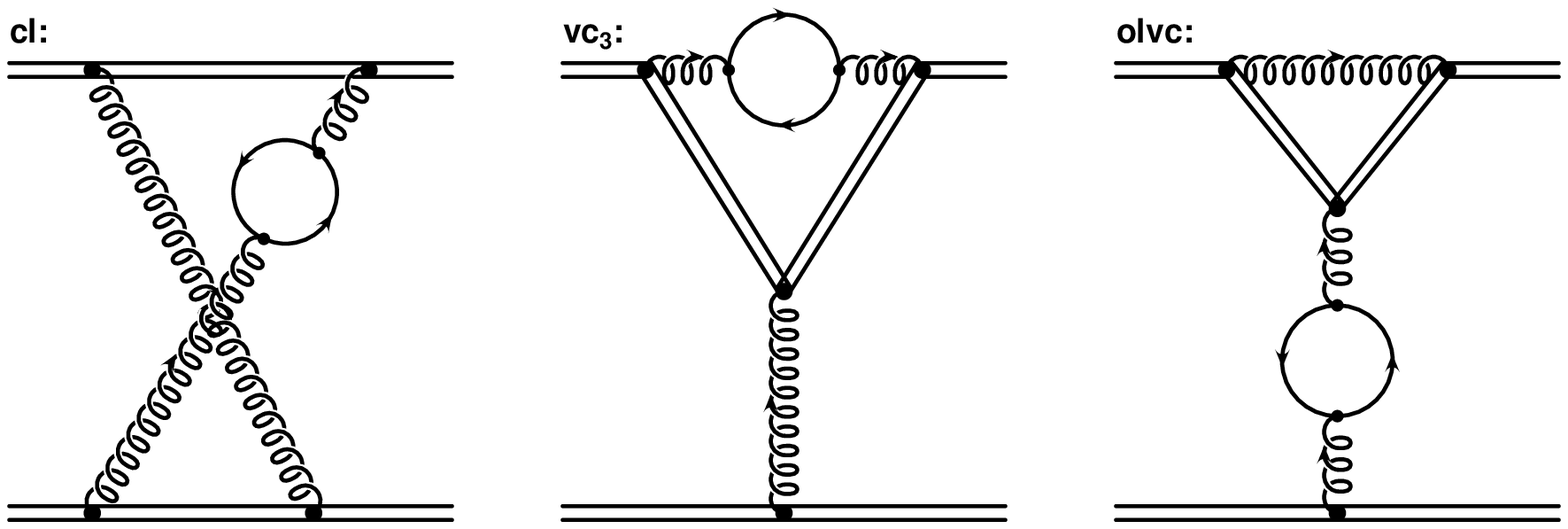,height=1.1in}
\vspace{0.5cm} \\
\epsfig{file=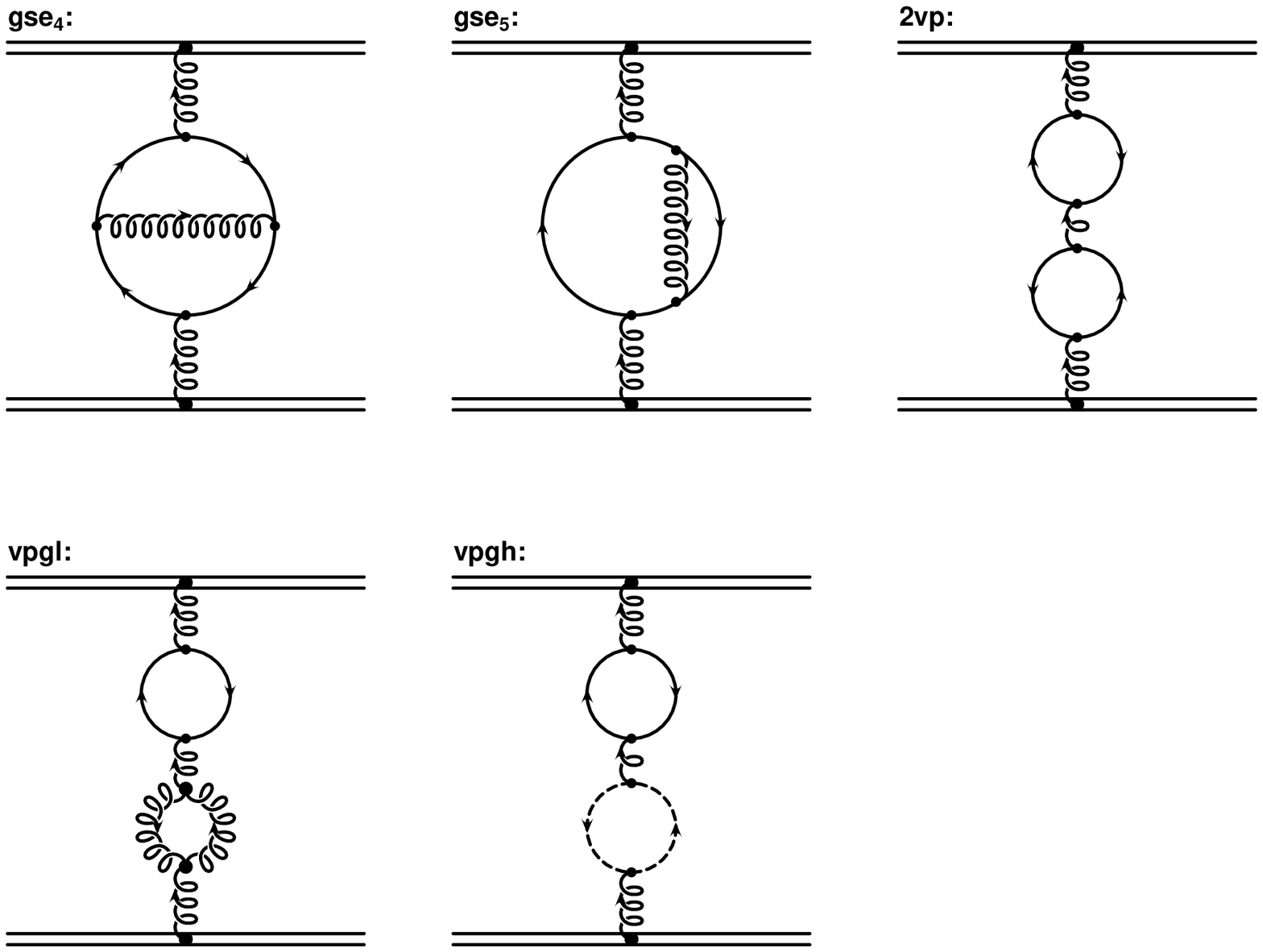,height=2.5in}
\vspace{0.5cm} \\
\caption{The two-loop massive fermionic corrections to the heavy quark
potential (from \protect\cite{melles98}). Double lines denote the heavy
quarks, single
lines the ``light'' quarks with mass $m$. The first two rows contain diagrams
with a typical non-Abelian topology. The middle line includes the infrared
divergent ``Abelian'' Feynman diagrams. They contribute to the potential only
in the non-Abelian theory due to color factors $\propto C_F C_A$. In addition,
although each diagram is infrared divergent, their sum is infrared
finite. The infrared finite Feynman diagrams with an Abelian topology
plus the diagrams consisting of one-loop
insertions with non-Abelian terms are shown in the last two rows.}
\label{fig:tlfd}
\end{figure}

The physical charge $\alpha_V(Q,m)$ can be expressed as a
perturbative series in any other renormalization scheme. For example, in the
minimal subtraction scheme, the perturbative series has the form:
\begin{eqnarray}
\alpha_V(Q,m) &=& \alpha_{{\mbox{\tiny MS}}}(\mu)
\left( 1 + v_1 (Q,m(\mu),\mu) \frac{
\alpha_{{\mbox{\tiny MS}}}(\mu)}{\pi} + v_2 (Q,m(\mu),\mu)
\frac{\alpha^2_{{\mbox{\tiny MS}}}(\mu)}{\pi^2} + \cdots \right)
\label{eq:aVmu}
\end{eqnarray}
where the massless limit of the 
coefficients $v_1$ and $v_2$ are known in the literature
\cite{Susskind,Fischler,Appelquist_Dine_Muzinich,Feinberg,Billoire,Peter,Schroder}.
Since the physical charge $\alpha_V(Q,m)$ cannot depend on the
renormalization scale $\mu$,
the $\mu$-dependence on the right-hand side of Eq.~(\ref{eq:aVmu})
must cancel to the order we are working.
Notice that the coefficients also depend on the renormalization scale $\mu$
used for the mass renormalization, i.e. through the dependence of
the running mass $m(\mu)$. Fig.~\ref{fig:tlfd} shows the Feynman diagrams
for the fermionic contributions to the two-loop coefficient $v_2(Q,m(\mu),\mu)$.
These contributions depend on the mass renormalization used for
the one-loop coefficient $v_1(Q,m(\mu),\mu)$. Since we are
predominantly interested in the  flavor-threshold dependence of heavy
quarks, we shall relate the running mass to the pole mass which is
renormalization-scale independent and gives explicit decoupling. This also
provides a physical picture as well as a straightforward Abelian
limit.

The next-to-leading order 
relation between the MS mass $m(\mu)$ and the pole mass $m$
is given by \cite{mspole},
\begin{equation}
m(\mu) = m \left[ 1 - C_F \frac{\alpha_{{\mbox{\tiny MS}}}(\mu)}{\pi}
\left( 1 + \frac{3}{2} \log \frac{
\mu}{m} -
\frac{3}{4}\left[\gamma - \log ( 4 \pi)\right] \right) \right],
\label{eq:mrm}
\end{equation}
where $\gamma$ is the Euler constant.
Inserting Eq.~(\ref{eq:mrm}) into  Eq.~(\ref{eq:aVmu}) gives at 
next-to-next-to-leading order
\begin{eqnarray}
\alpha_V(Q,m) &=& \alpha_{{\mbox{\tiny MS}}}(\mu)
\left[ 1 +
v_1 (Q,m,\mu) \frac{\alpha_{{\mbox{\tiny MS}}}(\mu)}{\pi} +
\left[v_2 (Q,m,\mu) + \Delta_m(Q,m,\mu)\right]
\frac{\alpha_{\mbox{\tiny MS}}^2(\mu)}{\pi^2}
\right] \label{eq:aVmupm}
\end{eqnarray}
where $\Delta_m(Q,m,\mu)$ denotes the contribution arising from $v_1$
when changing from the MS mass to the pole mass:
$v_1(Q,m(\mu),\mu) = v_1(Q,m,\mu) + \Delta_m(Q,m,\mu)
\frac{\alpha_{{\mbox{\tiny MS}}}(\mu)}{\pi}.$

The Gell-Mann Low function \cite{gl} for the $V$-scheme is defined as the
total logarithmic 
derivative of the effective charge with respect to the physical
momentum transfer scale $Q$:
\begin{equation}\label{eq:psiv}
\Psi_V \left( \frac{Q}{m} \right) \equiv  \frac{d \alpha_V (Q,m)}{
d \log Q} \equiv \sum^{\infty}_{i=0} -\psi_{V}^{(i)} \frac{\alpha_V^{i+2} (Q,m)}
{\pi^{i+1}} \; , 
\end{equation}
where in the massless case the coefficients $\psi_{V}^{(0)}$ and
$\psi_{V}^{(1)}$ are given by,
\begin{eqnarray}
\psi_{V}^{(0)}\left(m=0 \right) &=&  \frac{11}{6}N_C-\frac{1}{3}N_F =
\frac{11}{2}-\frac{1}{3}N_F \; ,\\
\psi_{V}^{(1)} \left( m=0 \right)&=&
\frac{17}{12}N_C^2-\frac{5}{12}N_CN_F-\frac{1}{4}C_FN_F
= \frac{51}{4} -\frac{19}{12}N_F \; .
\end{eqnarray}
For the massive case all the mass effects will be collected into a
mass-dependent function $N_F$. In other words we will write
\begin{eqnarray}\label{eq:nfv0}
\psi_{V}^{(0)}\left( \frac{Q}{m}  \right) &=&
\frac{11}{2}-\frac{1}{3}N_{F,V}^{(0)}\left( \frac{Q}{m} \right) \\
\label{eq:nfv1}
\psi_{V}^{(1)} \left(  \frac{Q}{m}  \right)&=&
 \frac{51}{4} -\frac{19}{12}N_{F,V}^{(1)}\left( \frac{Q}{m} \right) \; ,
\end{eqnarray}
where the subscript $V$ indicates the scheme dependence of
$N_{F,V}^{(0)}$ and $N_{F,V}^{(1)}$.

Taking the derivative of  Eq.~(\ref{eq:aVmupm}) with respect to $\log Q$ and
re-expanding the result in  $\alpha_V(Q,m)$ gives the following equations
for the first two coefficients of $\Psi_V$:
\begin{eqnarray}
\psi_{V}^{(0)}\left( \frac{Q}{m} \right) &=& -\frac{d v_1 (Q,m,\mu)}
{d \log Q} \label{eq:psi0} \\
\psi_{V}^{(1)} \left( \frac{Q}{m} \right)&=&
-\frac{d [v_2 (Q,m,\mu)+ \Delta_m(Q,m,\mu)]}{d \log Q} + 2
v_1 (Q,m,\mu) \frac{d v_1 (Q,m,\mu)}{d \log Q} \; .
\label{eq:psi1}
\end{eqnarray}
The argument $Q/m$  indicates that there is no
renormalization-scale  dependence in Eqs.~(\ref{eq:psi0}) and
(\ref{eq:psi1}). Rather, $\psi_{V}^{(0)}$ and $\psi_{V}^{(1)}$ are functions
of the ratio of the physical momentum transfer
$Q = \sqrt{-q^2}$ and the pole mass $m$ only.
The expression for $\psi_{V}^{(0)}$ agrees
with our result in Ref.~\cite{bgmr}.
In Eq.~(\ref{eq:psi1}) the derivative of the $\Delta_m(Q,m,\mu)$-term comes
from using the pole-mass instead of the MS mass, whereas
the remaining mass dependence in  Eq.~(\ref{eq:psi1})  is arbitrary
in the sense that a different mass scheme is formally of higher order.
In addition we note that
the contribution $2v_1 dv_1/d \log Q$ cancels the reducible
contribution (labeled  {\bf 2vp} in Fig.~\ref{fig:tlfd}) to
$v_2$;  it is thus sufficient to consider one quark flavor
at a time.

\section{Numerical Results for the Analytic $N_F$}\label{sec:nr}

Because of the complexity of the integrals encountered in the
evaluation~\cite{melles98} of the massive two-loop corrections to the
heavy quark potential, the  results were obtained
numerically using the adoptive Monte Carlo integration program
VEGAS~\cite{vegas}. Thus the derivative of the two-loop term
$v_2$ was calculated numerically, whereas the other terms in
Eqs.~(\ref{eq:psi0}) and (\ref{eq:psi1}) were obtained analytically.
The results are given in terms
of the contribution to the effective number of flavors 
$N^{(0)}_{F,V}\left( \frac{Q}{m} \right)$ and  
$N^{(1)}_{F,V}\left( \frac{Q}{m} \right)$ in the $V$-scheme, 
from a given quark with mass $m$, 
defined according to Eqs.~(\ref{eq:nfv0}) and (\ref{eq:nfv1})
respectively. The Appelquist-Carazzone~\cite{ac}  theorem
requires the decoupling of heavy masses at small momentum
transfer for physical observables. Thus we expect 
$N^{(1)}_{F,V} \left(
\frac{Q}{m} \right)$ to go to zero for  $Q/m \to 0$.  The massless result
$N_{F,V}^{(1)} \to 1$  must also be recovered for large scales.

The calculation presented in Ref.~\cite{melles98} required the evaluation
of four-dimensional integrals over Feynman parameters.  
Our results are based on 50 iterations of the integration grid each comprising 
$10^7$ evaluations of the function which were needed to achieve 
adequate convergence.  Even so,
the Monte Carlo results still are not completely stable for  small
values of $Q/m$, especially in the light of the numerical differentiation
required in Eq.~(\ref{eq:psi1}).  Nevertheless,
accurate results can be obtained by fitting the numerical
calculation to a suitable analytic function as shown below.

The one-loop contribution to the effective number of flavors $N_F$
follows from the standard  formula for QED vacuum polarization.
In our earlier paper~\cite{bgmr}
we used the simple representation in terms of a rational 
polynomial~\cite{derujula}:
\begin{equation}
N^{(0)}_{F,V} \left( \frac{Q}{m} \right) \approx \frac{1}{1+5 \frac{m^2}{Q^2}}
= \frac{0.2\frac{Q^2}{m^2}}{1+0.2 \frac{Q^2}{m^2}}
\label{eq:psi0fit}
\end{equation}
which displays decoupling  for small
scales and the correct massless limit at large
scales.  Similarly,  the numerical results for the two-loop contribution
can be fit to the form
\begin{equation}
N^{(1)}_{F,V} \left( \frac{Q}{m} \right) \approx
\frac{a_1 \displaystyle\frac{Q^2}{m^2}+a_2 \displaystyle\frac{Q^4}{m^4}}
{1+a_3 \displaystyle\frac{Q^2}{m^2} + a_2\displaystyle\frac{Q^4}{m^4}}
\label{eq:psi1fit}
\end{equation}
The parameter values $a_i$ and the errors obtained from the fit to the
numerical calculation in the $V$-scheme for QCD and QED are given in Table
\ref{tab:fit}. Similar decoupling forms have been used for
interpolating the flavor dependence of the effective coupling in the
momentum subtraction schemes (MOM)~\cite{yh,jt}.

\begin{table}[tb]
\caption{ Values for the parameters $a_1$, $a_2$ and $a_3$ in the
$V$-scheme for QCD and QED obtained by fitting our numerical results to
the form given by Eq.~(\ref{eq:psi1fit}). The $\chi^2$ values
were obtained by subscribing a constant $0.01$ error to each data point
in the fit.  (Here $d.o.f.$ denotes the effective number of degrees of
freedom for the fit, i.e. the number of fitted points minus the number of
parameters).}\
\label{tab:fit}
\begin{tabular}{lcccc}
     &   $a_1$            &   $a_2$            &  $a_3$            &
$\frac{\chi^2}{d.o.f.}$ \\
QCD  & $-0.571 \pm 0.034$ &  $0.221 \pm 0.015$ & $1.326 \pm 0.116$ &
$\frac{86}{43}$ \\
QED  & $1.069 \pm 0.0088$ & $0.0133 \pm 0.0002$ & $0.402 \pm 0.005$ &
$\frac{121}{46}$
\end{tabular}
\end{table}

In the case of QCD we obtain the following approximate form for the
effective number of flavors for a given quark with mass $m$:
\begin{equation}
N^{(1)}_{F,V} \left( \frac{Q}{m} \right) \approx \frac{\left( -0.571 + 0.221
\displaystyle\frac{Q^2}{m^2} \right)
\displaystyle\frac{Q^2}{m^2}}{1+1.326 \displaystyle\frac{Q^2}{m^2} + 0.221
\displaystyle\frac{Q^4}{m^4}}
\label{eq:psi1qcd}
\end{equation}
and for QED
\begin{equation}
N^{(1)}_{F,V} \left( \frac{Q}{m} \right) \approx \frac{\left( 1.069 + 0.0133
\displaystyle\frac{Q^2}{m^2} \right)
\displaystyle\frac{Q^2}{m^2}}{1+0.402 \displaystyle\frac{Q^2}{m^2} + 0.0133
\displaystyle\frac{Q^4}{m^4}} \; .
\label{eq:psi1qed}
\end{equation}

\begin{figure}[htbp]
\center
\epsfig{file=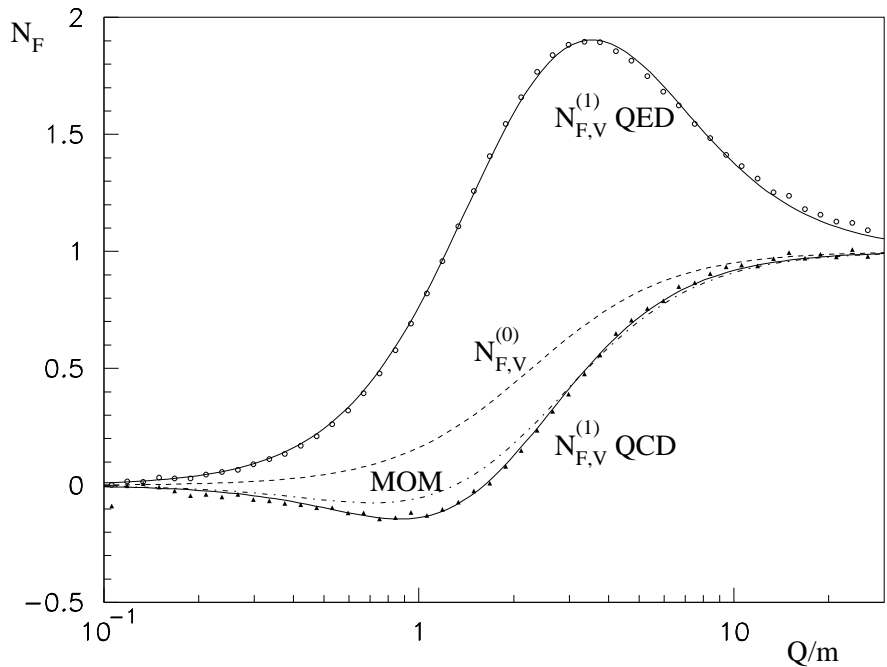,width=12cm}
\caption{The numerical results for the gauge-invariant  $N_{F,V}^{(1)}$ in QED
(open circles) and QCD (triangles) with the best $\chi^2$ fits of
Eqs.~(\ref{eq:psi1qed}) and (\ref{eq:psi1qcd})
superimposed respectively.   The dashed
line shows the one-loop $N_{F,V}^{(0)}$ function  of Eq.~(\ref{eq:nf0}).
For comparison we
also show the gauge dependent two-loop result obtained in MOM schemes
(dash-dot)  \protect\cite{yh,jt}. At large $\frac{Q}{m}$ the theory becomes
effectively massless, and both schemes agree as expected. The figure also
illustrates the decoupling of heavy quarks at small $\frac{Q}{m}$.}
\label{fig:nfV}
\end{figure}

The results of our numerical calculation of $N_{F,V}^{(1)}$ in the
$V$-scheme for QCD and QED are shown in Fig.~\ref{fig:nfV}.
The decoupling of heavy quarks becomes manifest at small $Q/m$, and
the massless limit is attained for large $Q/m$. The QCD form actually 
becomes negative at moderate values of $Q/m$, a novel feature of
the anti-screening non-Abelian contributions. This property is
also present in the (gauge dependent) MOM results.
In contrast, in Abelian QED the two-loop contribution to
the effective number of flavors becomes larger than 1 at intermediate
values of $Q/m$.  We also display the one-loop
contribution $N^{(0)}_{F,V} \left( \frac{Q}{m} \right)$ which
monotonically  interpolates between the decoupling and massless limits.
The solid curves displayed in
Fig.~\ref{fig:nfV} shows that the parameterizations of
Eq.~(\ref{eq:psi1qcd})  which we used for fitting the numerical results are
quite accurate. This is also indicated by the 
$\chi^2$ values obtained for the fits as given in Table
\ref{tab:fit}.

\begin{figure}[htbp]
\center
\epsfig{file=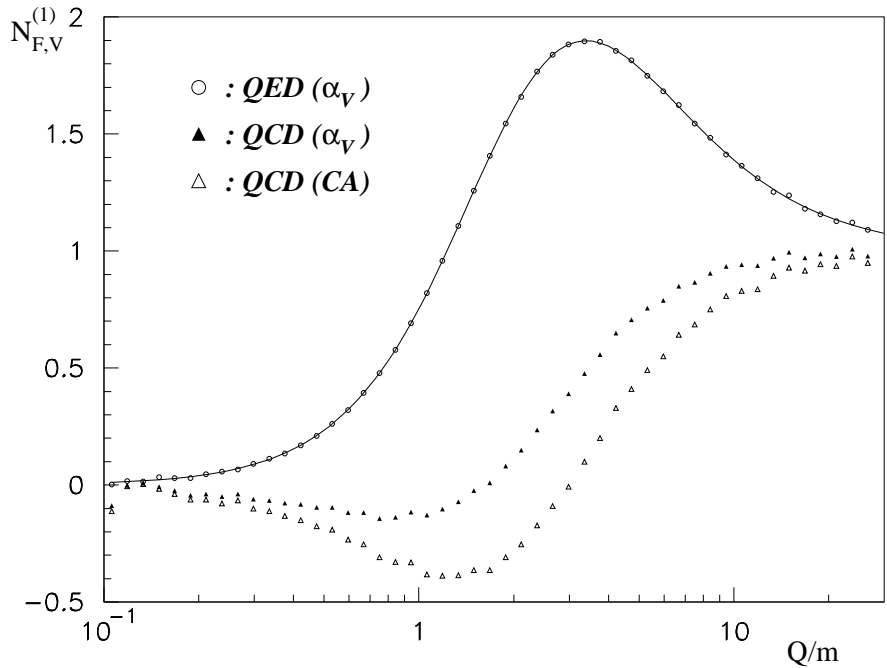,width=12cm}
\caption{Comparison of the Abelian limit of our results (open circles)
for $N_{F,V}^{(1)}$  based on the calculation in 
Ref.~\protect\cite{melles98} which was done 
in the MS-scheme with the well known result in
the literature \protect\cite{ks,br,k} done in the on-shell renormalization
scheme (solid line).  Also shown are the gauge invariant non-Abelian
contribution only ($\propto C_A$) (open triangles) as well as the sum of all
terms in QCD (solid triangles). The correct Abelian behavior is a very strong
check on the results given in Ref.~\protect\cite{melles98}. All Monte Carlo
results are based on $10^7$ evaluations per iteration and 50 iterations
of the integration grid.}
\label{fig:checks}
\end{figure}

A strong check of our results, as well as the results presented in 
Ref.~\cite{melles98}, is the agreement with the two-loop Gell-Mann
Low function for
QED~\cite{ks,br,k}.  Fig.~\ref{fig:checks} contains detailed comparisons of
the analytic QED result with our numerical computation. For comparison the
figure also displays the purely non-Abelian  part of the QCD result as well
as the total QCD result.  The scalar functions occurring in the Abelian
corrections are also used in the evaluation of the non-Abelian  contributions
($\propto C_A$), and it is therefore important to know that 
they were  calculated correctly.

\begin{figure}[t]
\center
\epsfig{file=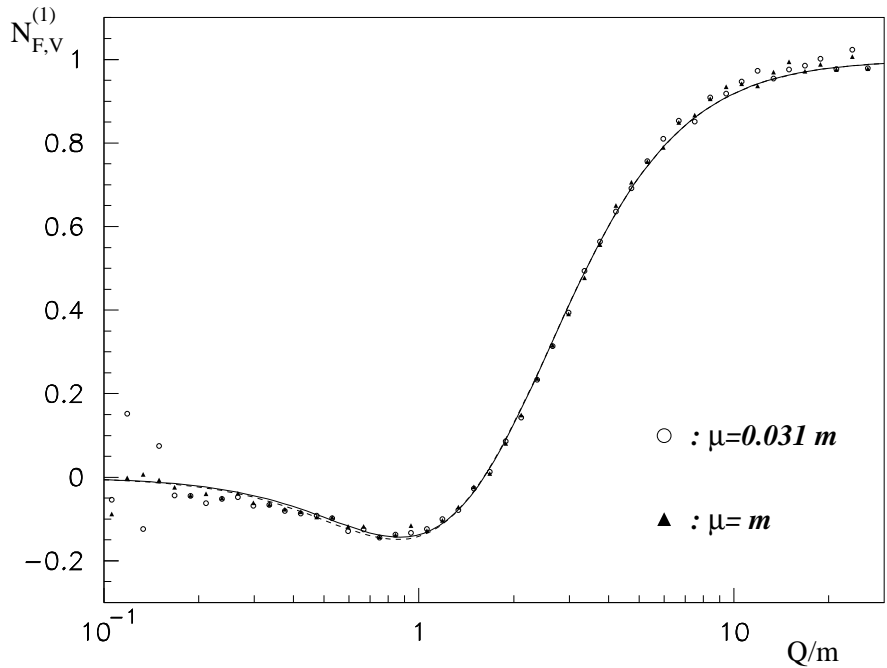,width=12cm}
\caption{Illustration of the renormalization scale independence of
the two-loop effective number of flavors $N_{F,V}^{(1)}$  as a function of the
ratio of the physical momentum transfer $Q$ over the pole mass $m$.  Numerical
instabilities  are visible for small values of $\frac{Q}{m}$ and occur because
of limited Monte Carlo statistics ($10^7$ evaluations for each
of the 50 iterations).  The two fits obtained, which agree within statistical
errors, are shown as a solid and dashed line for $\mu=m$ and $\mu=0.031m$
respectively.}
\label{fig:muind}
\end{figure}

Another important test of our results is
renormalization-scale ($\mu$) independence, which follows from the
fact that the effective number of flavors  in the $V$-scheme is a physical
quantity.
This is illustrated in Fig.~\ref{fig:muind} which shows the
results obtained for two different renormalization scales
($\mu=0.031m$ and $\mu=m$). The figure also shows the fits obtained for
the two different cases. In fact, the differences are so small that the two
lines cannot easily be distinguished.

We can also apply  the same fitting procedure  to the dependence of
the one-loop effective
$N_F$:
\begin{equation}
N^{(0)}_{F,V} = \frac{1}{1+\left(5.19 \pm 0.03 \right) \frac{m^2}{Q^2}}
 \;\;\;;\; \frac{\chi^2}{d.o.f.} = \frac{19}{27} \label{eq:nf0}
\end{equation}
This gives a higher precision global fit compared to the form
in Eq.~(\ref{eq:psi0fit}).

\section{Some properties of the analytic 
coupling in the V-scheme}\label{sec:prop} 

Using the numerical results for  $N^{(0)}_{F,V}$ and $N^{(1)}_{F,V}$ 
the evolution equation (\ref{eq:psiv}) can be solved numerically using the 
classical Runge-Kutta algorithm. As starting value we
use $\alpha_V(M_Z,m) = 0.126$ in next-to-leading order and 
$\alpha_V(M_Z,m) = 0.134$ in leading order which have been obtained
from the  value $\alpha_{\overline{\mbox{\tiny MS}}}(M_Z)=0.118$.
It should be noted that it is straight forward to solve
this equation numerically
since we are using the pole-masses which do not depend on $Q$.
This should be compared with the MOM-scheme where one gets two coupled
differential equations to solve, both for the coupling and the mass.

\begin{figure}[htb]
\begin{center}
\mbox{\epsfig{figure=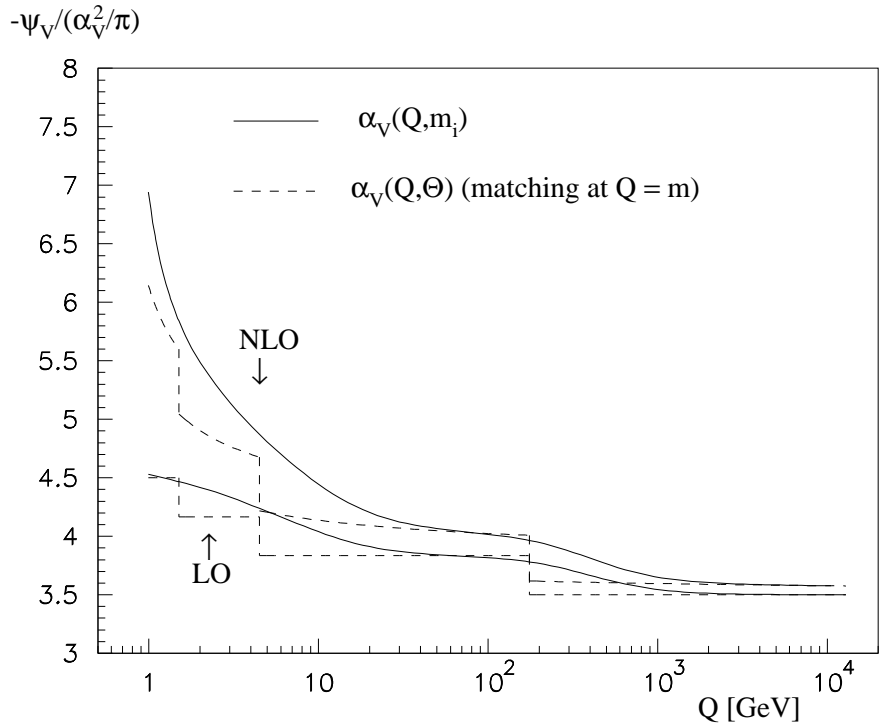,width=12cm}}
\end{center}
\caption[*]{The  scaled $\Psi$-function, 
$- {\Psi}_{V}/({\alpha}^2_{V}/\pi)$ in the analytic V-scheme 
${\alpha}_{V}(Q,m_i)$ (solid)
compared to the $\alpha_{V}(Q,\Theta)$  scheme with
discrete theta-function treatment of flavor
thresholds with continuous matching at $Q=m$ (dashed).}
\label{fig:psiv}
\end{figure}

The resulting leading and next-to-leading order $\Psi$-function
in the V-scheme is shown in Fig.~\ref{fig:psiv} scaled with
the leading dependence on ${\alpha}_{V}$, {\it i.e.}
$- {\Psi}_{V}/({\alpha}^2_{V}/\pi)$. For comparison
the figure also shows  the $\Psi$-function obtained with
discrete theta-function thresholds with continuous 
matching, ${\alpha}_{V}(Q,N_F=\Theta)={\alpha}_{V}(Q,N_F=\Theta+1)$, at the
naive matching scale $Q=m$.
As can be seen from the figure there are significant differences
between the two approaches both in leading and next-to-leading order.
In fact the difference becomes larger when going to next-to-leading order.
We also note that the scale dependence of the coupling is larger in 
next-to-leading order and that the convergence of the $\Psi$-function
is not very good for scales below a few GeV.
From the figure it is also clear that there are no plateaus in the analytic 
treatment of quark masses. Thus there is no region of the scale $Q$
below $\sim 1$ TeV  where all quark masses can be neglected at the same time.

\begin{figure}[htb]
\begin{center}
\mbox{\epsfig{figure=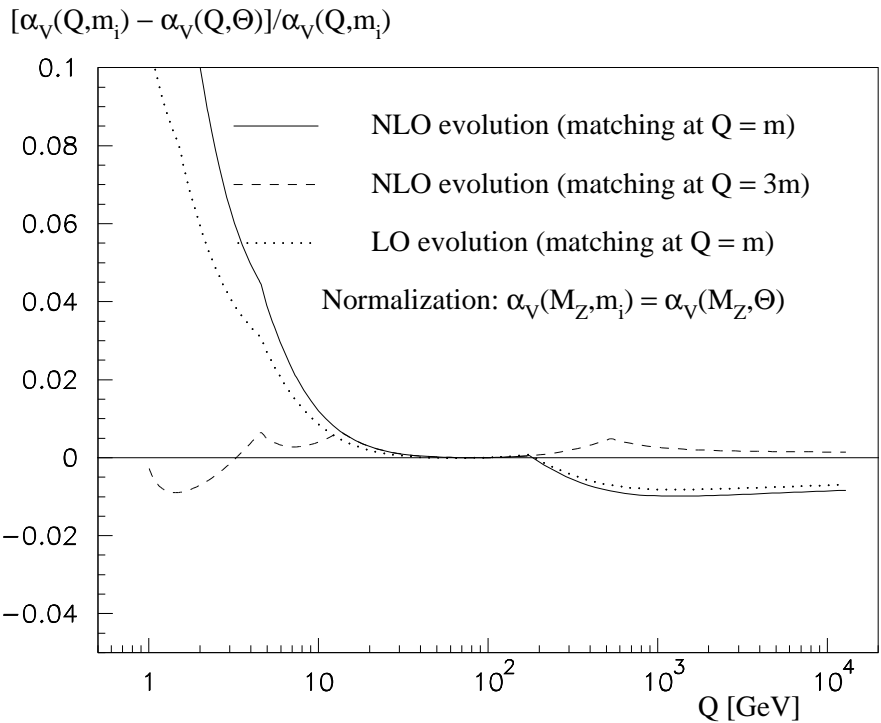,width=12cm}}
\end{center}
\caption[*]{The relative difference between the
solutions to
the evolution equation using the analytic
$\Psi$-function,
${\alpha}_{V}(Q,m_i)$, versus
discrete theta-function thresholds, 
$\alpha_{V}(Q,\Theta)$.
The solid (dashed) curve shows the next-to-leading (leading) order result.}
\label{fig:avdiffnlo}
\end{figure}

The solution of the evolution equation also gives 
the coupling as a function of the scale $Q$. 
The relative difference between the analytic
${\alpha}_{V}(Q,m_i)$ and 
 the  discrete theta-function treatment of flavor
thresholds with continuous matching at $Q=m$,
$\alpha_{V}(Q,\Theta)$, is shown in Fig.~\ref{fig:avdiffnlo} both 
in leading and next-to-leading order.

As can be seen from the figure, the difference between the
analytic and step-function treatment of quark masses in the running persists
when going to higher
order. In fact we expect this difference to remain
to all orders. The reason is that the $\Psi$ function is
not continuous in the step-function approach and the stepsize at the thresholds
is governed by the lowest order term $\psi^{(0)}$. Thus
there will always be a finite difference between the continuous
$\Psi$-function and the one with theta-function thresholds. 
The difference can be made smaller by modifying the matching scale
to be $Q=3m$ (but still using continuous matching) which is also illustrated in
the figure. However, the difference cannot be made smaller than $\sim 1\%$.
 The only way to include the finite quark mass effects in
the fixed flavor treatment is by making a higher twist
analysis to all orders in $m^2/Q^2$ and $Q^2/m^2$ for light and heavy quarks
respectively.

\begin{figure}[htb]
\begin{center}
\mbox{\epsfig{figure=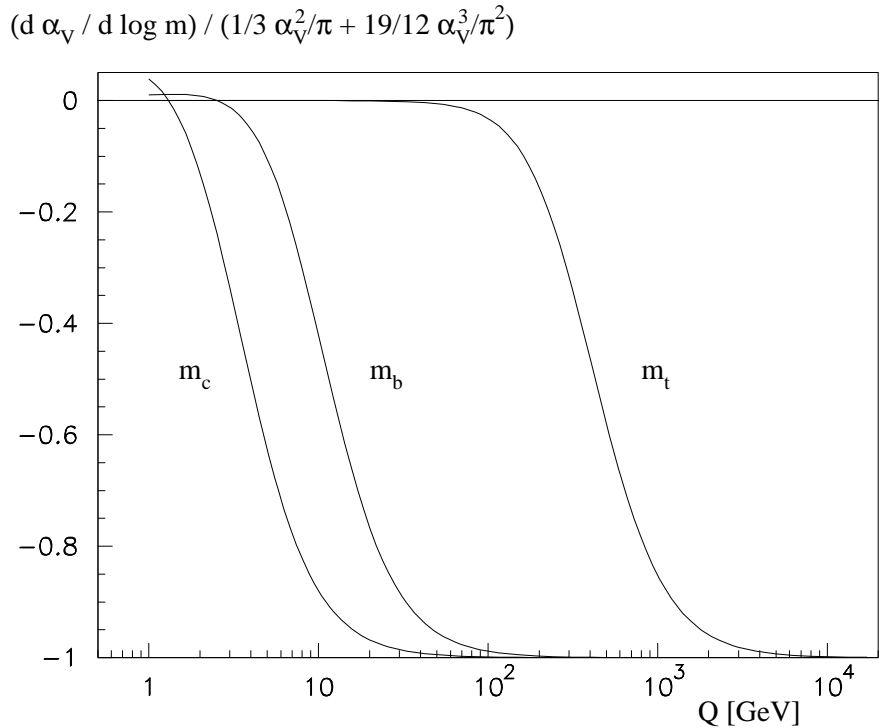,width=12cm}}
\end{center}
\caption[*]{The logarithmic derivative of $\alpha_V(Q,m_i)$ with respect to 
the quark mass for each of the heavy quarks. }
\label{fig:davdm}
\end{figure}

Noting that the differential equation for the scale dependence of the
coupling is homogeneous in $Q/m$ we can also get  
the logarithmic derivative of $\alpha_V(Q,m_i)$ with respect
to the heavy quark masses,
\begin{equation}
\frac{d \alpha_V(Q,m_i)}{d \log{m}}=- 
\left.\frac{d \alpha_V(Q,m_i)}{d \log{Q}}\right|_{q} \; ,
\end{equation}
where the subscript $q$ refers to the quark part of the $\Psi$-function.
The resulting mass-dependence is shown in Fig.~\ref{fig:davdm} 
for each of the heavy quarks.

The figure illustrates how the quarks decouple for small scales $Q$
and how they become effectively massless for large scales $Q$. 
In the intermediate 
region the mass-dependence depends on the ratio of the mass to the scale $Q$.
For large scales $Q$ the derivative approaches the asymptotic value
\begin{equation}
\left.\frac{d \alpha_V(Q,m_i)}{d \log{m}}\right|_{Q^2\gg m^2}=
-\frac{1}{3}\frac{\alpha_V^2(Q,m_i)}{\pi}
-\frac{5N_C+3C_F}{12}\frac{\alpha_V^3(Q,m_i)}{\pi^2}
\end{equation}
which is the same as the $N_F$ dependent part of the $\Psi$-function
apart from the differing sign. 
This can also be derived from the decoupling relations for matching
fixed $N_F$ couplings as for example is done in 
the $\overline{\mbox{MS}}$ scheme.

\section{ Application }\label{sec:appl}

The purpose of this section is to compare the treatment of finite
quark mass effects in the V-scheme with the standard treatment in 
the $\overline{\mbox{MS}}$ scheme. To do this comparison we will 
follow our earlier paper \cite{bgmr} and use the
non-singlet hadronic width of a Z-boson with arbitrary
mass $\sqrt{s}$ starting
from the physical mass $\sqrt{s}=M_Z$ for normalization.

\begin{figure}[htbp]
\begin{center}
\mbox{\epsfig{figure=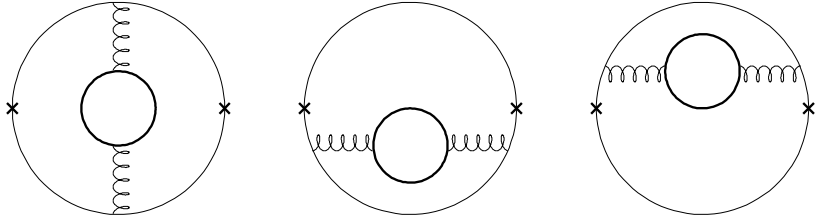,width=12cm}}
\end{center}
\caption[*]{The ``double bubble" diagrams. The crosses represent
the external electro weak current, the thin line is a massless quark
and the thick line is a massive quark.}
\label{fig:doublebubble}
\end{figure}

The finite quark mass effects that we are interested in are in  leading order
given by the ``double bubble" diagrams,  which are shown in
Fig.~\ref{fig:doublebubble}, where the outer quark loop which couples to the
weak current is considered massless and the inner quark loop is massive. These
corrections have been calculated in the $\overline{\mbox{MS}}$ scheme as
expansions in $m_q^2/s$ \cite{Hoang} and $s/m_Q^2$ \cite{Chetyrkin} for light
and heavy quarks, respectively,
  whereas they have been calculated numerically in
\cite{Soper_Surguladze}. In addition the $\alpha_{\mbox{\scriptsize{s}}}^3$
correction due to heavy quarks has been calculated as an expansion in $s/m_Q^2$
in \cite{L_R_V_NPB}. Other types of mass corrections, such as the 
double-triangle graphs where the external current is electroweak, 
are not taken into account.

The non-singlet hadronic width of
a Z-boson with arbitrary mass $\sqrt{s}$ is given by
\begin{equation}
\Gamma_{had}^{NS}(s)=\frac{G_Fs^{3/2}}{2\pi\sqrt{2}}
\sum_{q}\{(g_V^{q})^2+(g_A^{q})^2\}
\left[1+\frac{3}{4}C_F\frac{\alpha_{\Gamma}^{NS}(s)}{\pi}\right]  
\end{equation}
where $\alpha_{\Gamma}^{NS}(s)$ is the effective charge~\cite{grunberg} which
contains all QCD corrections.
In the following, the next-to-leading order expressions for the effective 
charge  $\alpha_{\Gamma}^{NS}(s)$ in the $\overline{\mbox{MS}}$ 
and V schemes will be compared for arbitrary $s$ using next-to-leading
order evolution starting from the physical mass $\sqrt{s}=M_Z$ which
is used as normalization condition.

\subsection{$\overline{\mbox{MS}}$ scheme treatment}

In the $\overline{\mbox{MS}}$ scheme the
effective charge  $\alpha_{\Gamma}^{NS}(s)$ is given by 
\begin{eqnarray} \label{eq:agmsfull}
\alpha_{\Gamma}^{NS}(s) & = &
\alpha_{\overline{\mbox{\tiny MS}}}^{(N_L)}(\mu)
+\left[r_{1,\overline{\mbox{\tiny MS}}}(\mu)+
\sum_{q=1}^{N_L}F_1\left(\frac{m_q^2}{s}\right)+
\sum_{Q=N_L+1}^{6}G_1\left(\frac{s}{m_Q^2}\right)\right]
\frac{\left(\alpha_{\overline{\mbox{\tiny MS}}}^{(N_L)}(\mu)\right)^2}{\pi} 
\nonumber \\ &&
+\left[r_{2,\overline{\mbox{\tiny MS}}}(\mu)+
\sum_{q=1}^{N_L}F_2\left(\frac{m_q^2}{s}\right)+
\sum_{Q=N_L+1}^{6}G_2\left(\frac{s}{m_Q^2}\right)\right]
\frac{\left(\alpha_{\overline{\mbox{\tiny MS}}}^{(N_L)}(\mu)\right)^3}{\pi^2}
\end{eqnarray}
where  the coefficients $r_1$ and $r_2$ are given by,
\begin{eqnarray*}
r_{1,\overline{\mbox{\tiny MS}}}(\mu=\sqrt{s})
& = & -\frac{1}{8}C_F+\frac{1}{12}N_C+
\left(\frac{11}{4}-2\zeta_3\right)\beta_0 
\nonumber \\ & = & 1.986-0.115N_F
\nonumber \\
r_{2,\overline{\mbox{\tiny MS}}}(\mu=\sqrt{s}) & = &
-\frac{23}{32}C_F^2+
\left(- \frac{53}{144}- \frac{11}{4}\zeta_3\right)N_C^2+
\left(-\frac{101}{192}+\frac{11}{4}\zeta_3\right)C_F N_C+
\nonumber \\ &&
\left(\frac{11}{4}-2\zeta_3\right)\beta_1+
\left(\frac{151}{18}-\frac{19}{3}\zeta_3-\frac{\pi^2}{12}\right) \beta_0^2 +
\nonumber \\ &&
\left[\left(-\frac{37}{32}-8\zeta_3+10\zeta_5\right)C_F+
\left(\frac{83}{48}+\frac{5}{6}\zeta_3-\frac{5}{3}\zeta_5\right)N_C
\right]\beta_0 
\nonumber \\ &=&  -6.637-1.200N_F-0.00518N_F^2
\end{eqnarray*}
(with $\beta_0=\psi_V^{(0)}(m=0)$ and $\beta_1=\psi_V^{(1)}(m=0)$) and 
the functions $F$ and $G$ are the effects of non-zero quark masses
for light and heavy quarks, respectively. The expansions of the
$\alpha_{\overline{\mbox{\tiny MS}}}^2$ finite quark mass corrections 
are given by\footnote{In our earlier paper~\cite{bgmr} 
there was a typographical
error giving the wrong signs for the two $\ln$-terms in $G_1$.}
\begin{eqnarray}
F_1\left(\frac{m^2}{s}\right) & = &
\left(\frac{m^2}{s}\right)^2\left[\frac{13}{3}
-4\zeta_3-\ln\left(\frac{m^2}{s}\right)\right]
\nonumber \\ &&
+ \left(\frac{m^2}{s}\right)^3\left[\frac{136}{243}+\frac{16}{27}\zeta_2
+\frac{56}{81}\ln\left(\frac{m^2}{s}\right)
-\frac{8}{27}\ln^2\left(\frac{m^2}{s}\right)\right]
\\
G_1\left(\frac{s}{m^2}\right) & = &
\frac{s}{m^2}\left[ \frac{44}{675}
-\frac{2}{135}\ln\left(\frac{s}{m^2}\right)\right]
+\left(\frac{s}{m^2}\right)^2\left[-\frac{1303}{1058400}
+\frac{1}{2520}\ln\left(\frac{s}{m^2}\right)\right]
\end{eqnarray}
which are accurate to within a few percent for $m_q^2/s<0.25$
and $s/m_Q^2<4$ respectively.
We will also use the relation \cite{Soper_Surguladze},
\begin{equation}
F\left(\frac{m^2}{s}\right) =
G\left(\frac{m^2}{s}\right) + \frac{1}{6}\ln\left(\frac{m^2}{s}\right)
-\left(-\frac{11}{12}+\frac{2}{3}\zeta_3 \right)
\end{equation}
to obtain $F_1$ in the interval $0.25 < m^2/s < 1$ where the expansion
of $F_1$ given above breaks down.

The $\alpha_{\overline{\mbox{\tiny MS}}}^3$ finite quark mass corrections 
are only known for heavy quarks ($G_2$), whereas the corresponding
corrections due to light quarks ($F_2$) have not yet been calculated. 
The known corrections in $G_2$ are small, for $m_Q^2=s$ they are of order
$G_2 \sim 0.1$ for the top quark.

The number of light flavors $N_L$ in the $\overline{\mbox{MS}}$ scheme
 is a function of the renormalization
scale $\mu$. In the following we will assume that the matching of the different
effective theories with different number of massless quarks is done at
the quark masses. In other words a quark with mass $m<\mu$ is considered
as light whereas a quark with mass  $m>\mu$ is considered as heavy.
In addition the $\overline{\mbox{MS}}$ quark masses are used. The general
matching condition  in the $\overline{\mbox{MS}}$ scheme is to next-to-leading
order given by~\cite{msmatch},
\begin{equation}
 \alpha_{\overline{\mbox{\tiny MS}}}^{(N_L)}(\mu)
= \alpha_{\overline{\mbox{\tiny MS}}}^{(N_L+1)}(\mu)-
\frac{1}{3}\log\left(\frac{\mu}{m(\mu)}\right)
\frac{\left(\alpha_{\overline{\mbox{\tiny MS}}}^{(N_L+1)}(\mu)\right)^2}{\pi} 
\end{equation}
where $m(\mu)$ is the mass of quark number $N_L+1$. 
The dependence on the matching 
scale can be made arbitrarily small by calculating the matching condition to 
high enough order. However this does not mean that the finite quark mass
effects are taken into account. The only way to include these mass effects in
the ordinary $\overline{\mbox{MS}}$ treatment is by making a higher twist
expansion to all orders in $m^2/Q^2$ and $Q^2/m^2$ for light and heavy quarks
respectively, {\it i.e.} the functions $F$ and $G$ given above.

In the following comparison 
we will restrict ourselves to the next-to-leading order
expression for  $\alpha_{\Gamma}^{NS}(s)$ in the $\overline{\mbox{MS}}$ 
scheme including the finite quark mass corrections, {\it i.e. }
\begin{eqnarray}\label{eq:agms}
\alpha_{\Gamma}^{NS}(s) & = &
\alpha_{\overline{\mbox{\tiny MS}}}^{(N_L)}(\mu)
+\left[r_{1,\overline{\mbox{\tiny MS}}}(\mu)+
\sum_{q=1}^{N_L}F_1\left(\frac{m_q^2}{s}\right)+
\sum_{Q=N_L+1}^{6}G_1\left(\frac{s}{m_Q^2}\right)\right]
\frac{\left(\alpha_{\overline{\mbox{\tiny MS}}}^{(N_L)}(\mu)\right)^2}{\pi}
\end{eqnarray}
with $\mu=\sqrt{s}$ and next-to-leading order matching done at the quark masses.

\subsection{V scheme treatment}

In order to  relate the hadronic width of the Z-boson to the $\alpha_V$ 
scheme we
will require the massless coefficients in the relation between 
$\alpha_{\overline{\mbox{\tiny MS}}}$ and $\alpha_V$ for $Q^2 \gg m^2$,
\begin{eqnarray}\label{eq:avinv}
\alpha_{\overline{\mbox{\tiny MS}}}(\mu)
 & = &
\alpha_V(Q) + c_{1,V}{\alpha_V^2(Q) \over \pi}+
c_{2,V}{\alpha_V^3(Q) \over \pi^2}+ \cdots \; ,
\end{eqnarray}
where the coefficients  are given by
\begin{eqnarray}
c_{1,V}
& = & \frac{2}{3}N_C+\left(-\frac{5}{6}+\ln\frac{Q}{\mu}\right)\psi_V^{(0)}
\nonumber \\
c_{2,V} & = &
\left(-\frac{5}{144}-\frac{16\pi^2-\pi^4}{64}+
\frac{11}{4}\zeta_3\right)N_C^2+
\left(\frac{385}{192}-\frac{11}{4}\zeta_3\right)C_FN_C
\nonumber \\ &&
+\left(-\frac{5}{6}+\ln\frac{Q}{\mu}\right)\psi_V^{(1)}
+\left(\frac{25}{36}-\ln^2\frac{Q}{\mu}\right)\left(\psi_V^{(0)}\right)^2 
\nonumber \\ &&
+\left[ 
\left(-\frac{35}{32}+\frac{3}{2}\zeta_3\right)C_F-
\left(\frac{103}{144}+\frac{7}{4}\zeta_3\right)N_C+
2c_{1,V}\ln\frac{Q}{\mu}\right]\psi_V^{(0)}.
\end{eqnarray}
The logarithmic $Q/\mu$ dependence of the coefficients 
 follows from requiring the expansion of $\alpha_V(Q)$ in 
 $\alpha_{\overline{\mbox{\tiny MS}}}(\mu)$
to be $\mu$ independent to the order we are working.
Inserting Eq.~(\ref{eq:avinv}) into the massless version of
Eq.~(\ref{eq:agmsfull}) ({\it i.e.} without the finite quark mass
corrections $F$ and $G$ for light and heavy quarks respectively) gives the
relation between the effective charges $\alpha_{\Gamma}^{NS} $ and $\alpha_V$
for $Q^2 \gg m^2$ which is independent of the
intermediate $\overline{\mbox{MS}}$ scheme:
\begin{eqnarray}
 \alpha_{\Gamma}^{NS}(s)  & = &
\alpha_V(Q) + r_{1,V}{\alpha_V^2(Q) \over \pi}+
r_{2,V}{\alpha_V^3(Q) \over \pi^2}+ \cdots \; .
\end{eqnarray}
where the coefficients are given by
\begin{eqnarray*}
r_{1,V}
& = & -\frac{1}{8}C_F-\frac{3}{4}N_C+
\left( \frac{23}{12}-2\zeta_3+\ln\frac{Q}{\sqrt{s}}\right)\psi_V^{(0)}
\nonumber \\
r_{2,V} & = &
-\frac{23}{32}C_F^2+
\frac{21}{16}C_F N_C+
\left(-\frac{16\pi^2-\pi^4}{64}-\frac{7}{24}\right)N_C^2
\nonumber \\ &&
+
\left( \frac{23}{12}-2\zeta_3+\ln\frac{Q}{\sqrt{s}}\right)\psi_V^{(1)} 
+\left(\frac{9}{2}-\frac{\pi^2}{12}-3\zeta_3
-\ln^2\frac{Q}{\sqrt{s}}\right)\left(\psi_V^{(0)}\right)^2 
\nonumber \\ &&
+\left[ 
\left(-\frac{49}{24}-\frac{13}{2}\zeta_3+10\zeta_5\right)C_F+
\left(\frac{109}{24}-\frac{43}{12}\zeta_3-\frac{5}{3}\zeta_5\right)N_C+
2r_{1,V}\ln\frac{Q}{\sqrt{s}}\right]\psi_V^{(0)} .
\end{eqnarray*}
It should  be noted that this way of writing the two-loop coefficient
$r_{2,V}$ in terms of $\psi_V^{(0)}$ and $\psi_V^{(1)}$ follows from
the conformal ansatz.

\begin{figure}[htb]
\begin{center}
\mbox{\epsfig{figure=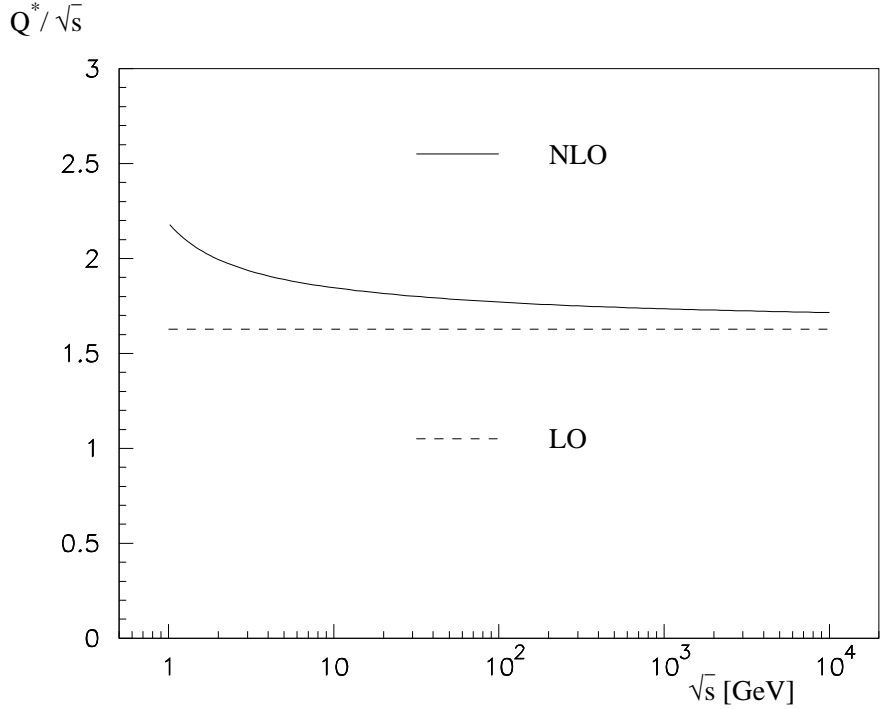,width=12cm}}
\end{center}
\caption[*]{The ratio of the commensurate scale $Q^*$ to $\sqrt{s}$
between the non-singlet width of the Z-boson and the heavy quark potential
as a function of $\sqrt{s}$ in next-to-leading (solid)
and leading (dashed) order.}
\label{fig:qratv}
\end{figure}

We now use the commensurate scale relation method 
to eliminate the scale ambiguity: 
$Q$ is set to $Q^*$ using the single scale scale-setting 
approach\cite{Kataev}, such that all
non-conformal terms proportional to
$\psi_V^{(0)}$ and $\psi_V^{(1)}$ are absorbed into the running of the
coupling~\footnote{There also exists a multiple scale setting 
approach~\cite{csr} where one has different scales for each order of 
$\alpha_V$. However, for clarity we concentrate on only one of the 
procedures. In addition, as noticed in our earlier paper~\cite{bgmr},
the multiple scale setting procedure does not always have the correct
Abelian limit.}.
This gives the next-to-leading order commensurate scale relation between
$\alpha_{\Gamma}^{NS} $ and $\alpha_V$. To obtain the next-to-next-to
leading order relation requires knowledge about the $N_F$-dependent part 
of the three-loop contribution.
Thus we arrive at the following commensurate scale relation between
$\alpha_{\Gamma}^{NS} $ and $\alpha_V$,
\begin{eqnarray}
\alpha_{\Gamma}^{NS}(\sqrt{s}) & = &
\alpha_V(Q^*) + 
\left(-\frac{1}{8}C_F+\frac{3}{4}N_C\right) {\alpha_V^2(Q^*) \over \pi}
\nonumber \\ &&
+\left[-\frac{23}{32}C_F^2+\frac{21}{16}C_F N_C+
\left(-\frac{16\pi^2-\pi^4}{64}-\frac{7}{24}\right)N_C^2
 \right]{\alpha_V^3(Q^*) \over \pi^2} 
\nonumber \\ & = &
\alpha_V(Q^*) + 
2.083 {\alpha_V^2(Q^*) \over \pi}-7.161 {\alpha_V^3(Q^*) \over \pi^2} .
\label{eq:agvfull}
\end{eqnarray}
The next-to-leading order commensurate scale $Q^*$ is given by
\begin{eqnarray}\label{eq:csr}
\frac{Q^*}{\sqrt{s}} & = & \exp\left\{-\frac{23}{12}+2\zeta_3
+\frac{\displaystyle
  \left[
  a_1 \psi_V^{(0)}(Q^*)+ a_2 \left(\psi_V^{(0)}(Q^*)\right)^2
  \right] \frac{\alpha_V(Q^*)}{\pi}  }
{\displaystyle\psi_V^{(0)}(Q^*) + 
 \psi_V^{(1)}(Q^*)\frac{\alpha_V(Q^*)}{\pi}}
\right\} 
\end{eqnarray}
where
\begin{eqnarray*}
a_1&=& \left(\frac{25}{16}-7\zeta_3-10\zeta_5\right)C_F
  +\left(-\frac{5}{3}+\frac{7}{12}\zeta_3+\frac{5}{3}\zeta_5\right)N_C 
  = 1.765\;,
 \\  
a_2&=& -\frac{119}{144}-\frac{14}{3}\zeta_3+4\zeta_3^2+\frac{\pi^2}{12} 
= 0.166 \;,
\end{eqnarray*}
which should be compared with the leading order commensurate scale
$Q^* = \sqrt{s}\exp\left(-\frac{23}{12}+2\zeta_3\right) = 1.628 \sqrt{s}$.
It should be noted that this way of writing the scale $Q^*$
differs slightly from the one used in \cite{Kataev}
in that it is written as the exponential
of a partial fraction where the denominator is proportional to the 
$\Psi$-function.
This ensures that the scale $Q^*$ has  sensible limits as $\alpha_V \to 0$ or
$\alpha_V \to \infty$ (for $N_F=3$, $Q^*=6.7Q$ in the limit
$\alpha_V \to \infty$).  If the coupling freezes or is bounded
for small scales, then the latter limit is of course not important. 
In addition $Q^*$ has the correct Abelian limit.
The resulting commensurate scale $Q^*$ is shown in Fig.~\ref{fig:qratv}
where it is also compared with the leading order scale. As can be seen
from the figure, the next-to-leading order correction to
the commensurate scale is small. 
The general convergence properties of the scale $Q^*$ as an
expansion in $\alpha_V$ is not known \cite{Mueller}.

The relation between $\alpha_{\Gamma}^{NS}(\sqrt{s})$ and 
$\alpha_V(Q^*)$ can be generalized   to be valid for all scales, 
for which perturbation theory is applicable, by
using the mass-dependent $\alpha_V(Q,m_i)$,
\begin{eqnarray}
\alpha_{\Gamma}^{NS}(\sqrt{s},m_i) & = &
\alpha_V(Q^*,m_i) + 
\left(-\frac{1}{8}C_F+\frac{3}{4}N_C\right) {\alpha_V^2(Q^*,m_i) \over \pi}
\nonumber \\ &&
 + \left[-\frac{23}{32}C_F^2+\frac{21}{16}C_F N_C+
\left(-\frac{16\pi^2-\pi^4}{64}-\frac{7}{24}\right)N_C^2
 \right]{\alpha_V^3(Q^*,m_i) \over \pi^2} ,
\end{eqnarray}
where the argument of $\alpha_{\Gamma}^{NS}$ is meant to indicated that the
quark mass effects related to the running of the coupling are taken into 
account and $m_i$ being the pole-masses for the quarks which do not depend 
on $Q$.
In addition we use the mass-dependent coupling $\alpha_V(Q,m_i)$
and the mass-dependent coefficients of the $\Psi_V$-function,
$\psi_V^{(0)}(Q,m_i)$ and $\psi_V^{(1)}(Q,m_i)$, in the
formula for $Q^{*}$ given by Eq.~(\ref{eq:csr}).
It should be noted that the scale $Q^*$ is only known to next-to-leading order.
Similarly the evolution equation for $\alpha_V(Q,m_i)$ is only known
to next-to-leading order. 
Therefore we can only consistently use the next-to-leading order result
when comparing with the treatment
of finite quark mass effects in the $\overline{\mbox{MS}}$ scheme, {\it i.e.}
\begin{eqnarray}\label{eq:agv}
\alpha_{\Gamma}^{NS}(\sqrt{s},m_i) & = &
\alpha_V(Q^*,m_i) + 
\left(-\frac{1}{8}C_F+\frac{3}{4}N_C\right) {\alpha_V^2(Q^*,m_i) \over \pi}  
\end{eqnarray}
where the scale $Q^*$ should be the leading order result for consistency.

\subsection{Comparison}

\begin{figure}[htbp]
\begin{center}
\mbox{\epsfig{figure=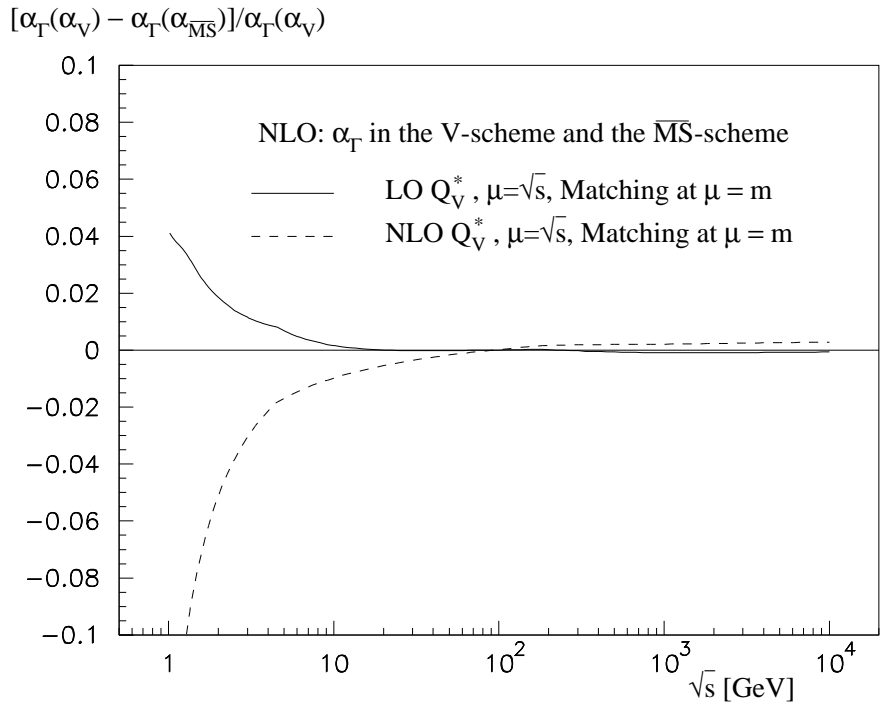,width=12cm}}
\end{center}
\caption[*]{ The relative difference between the
next-to-leading order expressions for $\alpha_{\Gamma}^{NS}$ 
in the $\overline{\mbox{MS}}$ and V schemes  respectively using next-to-leading
order evolution.}
\label{fig:agdiff}
\end{figure}

Fig.~\ref{fig:agdiff} shows the relative difference between the
next-to-leading order expressions for $\alpha_{\Gamma}^{NS}$ 
in the $\overline{\mbox{MS}}$ and V schemes given by 
Eqs.~(\ref{eq:agms}) and (\ref{eq:agv}) respectively. 
The predictions for the width have been 
normalized to the same value at $\sqrt{s}=M_Z$ 
using $\alpha_{\overline{\mbox{\tiny MS}}}^{(5)}(M_Z)=0.118$ and then
evolved using next-to-leading order evolution in the respective schemes.

The comparison shown in Fig.~\ref{fig:agdiff} illustrates
the relative difference between the predictions for 
$\alpha_{\Gamma}^{NS}$ in the $\overline{\mbox{MS}}$ and V schemes.
In our earlier paper \cite{bgmr} we showed that the different ways
of including the finite quark mass effects is smaller than $\sim 0.1\%$
by comparing the 
$\overline{\mbox{MS}}$  scheme with the analytic extension of the 
same which properly takes into account the 
flavor threshold effects analytically.
Therefore the difference between the $\overline{\mbox{MS}}$ and V scheme
predictions for $\alpha_{\Gamma}^{NS}$ can be attributed to the 
scheme dependence. This is illustrated by the fact that when using the 
next-to-leading order approximation for the commensurate scale, 
instead of the leading order one, the relative difference
changes sign and even becomes larger. This sensitivity is a consequence 
of the scale dependence of the coupling, especially at small
scales where the $\Psi$-function is large.
The proper inclusion of the finite quark mass effects is verified
by the smoothness of the curve.

\section{Summary and Conclusions}\label{sec:sum}
We have presented the calculation of $\psi_V^{(1)}$, the two-loop
term in the Gell-Mann Low function for the $\alpha_V$ scheme,
with massive quarks. This gives for the first time a gauge invariant
and renormalization scheme independent two-loop result for the
effects of quark masses in the running of the coupling. Renormalization
scheme independence is achieved by using the pole mass definition for
the ``light" quarks which contribute to the scale dependence of
the static heavy quark potential. Thus the pole mass and the $V$-scheme
are closely connected and have to be used in conjunction to give
reasonable results. The results of the calculation are presented in
numerical form due to the complexity of the integrations required. An important
cross-check is the successful reproduction of the well-known QED results.

The effective number of flavors in the two-loop coefficient of the 
Gell-Mann Low function in the $\alpha_V$ scheme, $N_{F,V}^{(1)}$,
becomes slightly negative for intermediate values of $Q/m$. 
This feature can be
understood as anti-screening from the non-Abelian contributions and
should be contrasted with the QED case where the effective number of
flavors becomes larger than one for intermediate $Q/m$. For small
$Q/m$ the heavy quarks decouple explicitly as expected in a physical
scheme, and for large $Q/m$ the massless result is retained.

The analyticity of the $\alpha_V$ coupling can be utilized to
obtain predictions for perturbatively calculable  observables
including the
finite quark mass effects associated with the running of the
coupling. By employing the commensurate scale relation method, 
observables which have been calculated in the
$\overline{\mbox{MS}}$ scheme can be related to the analytic V-scheme
without any scale ambiguity. The commensurate scale relations
provides the relation between the physical scales of two effective charges
where they pass through a common flavor threshold.

As an example, we have shown how to calculate the finite quark mass
corrections connected with the running of the coupling for
the non-singlet hadronic width of the Z-boson compared
with the standard treatment in the $\overline{\mbox{MS}}$ scheme.
The analytic treatment in the V-scheme gives a simple and straightforward
way of incorporating these effects for any observable. This should
be contrasted with the $\overline{\mbox{MS}}$ scheme where higher
twist corrections due to finite quark mass threshold effects have to be 
calculated separately for each observable.
The V-scheme is especially 
suitable for problems where the quark masses are important such as for
threshold production of heavy quarks and the hadronic width of the
$\tau$ lepton.

\acknowledgements
J.~R.~would like to thank the Physics department at Durham for their kind
hospitality during a visit when parts of this work were done.

This work was
supported in part by the EU Fourth Framework Programme
`Training and
Mobility of Researchers', Network `Quantum Chromodynamics
and the Deep
Structure of Elementary Particles', contract FMRX-CT98-0194
(DG 12 -
MIHT).

\newpage
\appendix
\section{Analytic $\alpha_{\overline{\mbox{\tiny{MS}}}}$ at Two
Loops}

Finite quark masses
are included naturally into the  running of  $\alpha_V$, thus
providing an analytic definition of the gauge theory coupling.
Furthermore, there is no scale  ambiguity in $\alpha_V(Q)$ since
the argument of the coupling is by definition the physical momentum transfer
$Q$. These advantages can be
carried over to the ordinary $\overline{\mbox{MS}}$  scheme by relating it
to the
physical
$\alpha_V$ scheme via a commensurate scale relation connecting
the two schemes.
It is in fact possible to combine the computational advantages of the
$\overline{\mbox{MS}}$ scheme and the physical and analytic
properties $\alpha_V$ scheme into one common scheme, the analytic extension of
the
$\overline{\mbox{MS}}$ scheme \cite{bgmr}.
However, as already mentioned in the introduction,
the conformal coefficients in the commensurate scale
relation between the $\alpha_V$ and
$\overline{\mbox{MS}}$ schemes does not preserve one of the
 defining criterion 
of the potential expressed in the bare charge, namely the non-occurrence of
color factors corresponding to an iteration of the potential. This
is probably an effect of the breaking of conformal invariance by
the $\overline{\mbox{MS}}$ scheme. The breaking of conformal symmetry
has also been observed when dimensional regularization is used as a
factorization scheme in both exclusive\cite{Frishman,Muller} and
inclusive\cite{Blumlein} reactions.  Thus, it does not turn out to be
possible to extend the modified scheme
${\widetilde \alpha}_{\overline{\mbox{\tiny MS}}}$ beyond leading order
without running into an intrinsic contradiction with conformal
symmetry.   
For completeness we
give the results of such an extension in this appendix.

\subsection{Commensurate scale relation between \mbox{$\alpha_{V}$} and
\mbox{$\alpha_{\overline{\mbox{\tiny MS}}}$} }

Our starting point for relating the \mbox{$\alpha_{V}$} and
\mbox{$\alpha_{\overline{\mbox{\tiny MS}}}$} schemes is the massless result for
$Q^2 \gg m^2$ which is given by Eq.~(\ref{eq:avinv}). Just as before we use the
commensurate scale relation method to eliminate  the scale ambiguity: the
scale  $Q$  is  set to $Q^*$ using the single scale scale-setting 
approach\cite{Kataev}, such that all nonconformal terms proportional to
$\psi_V^{(0)}$ and $\psi_V^{(1)}$ are absorbed into the running of the
coupling.   This gives the following commensurate scale relation between
$\alpha_{\overline{\mbox{\tiny MS}}}$ and $\alpha_V$,
\begin{eqnarray}\label{eq:csrmsofv}
\alpha_{\overline{\mbox{\tiny MS}}}(Q)
 & = &
\alpha_V(Q^{*}) +  \frac{2}{3}N_C{\alpha_V^2(Q^{*}) \over \pi}
\nonumber \\ &  & +
\left[\left(-\frac{5}{144}-\frac{16\pi^2-\pi^4}{64}+
\frac{11}{4}\zeta_3\right)N_C^2+
\left(\frac{385}{192}-\frac{11}{4}\zeta_3\right)C_FN_C\right]
{\alpha_V^3 \over \pi^2}
\nonumber \\ & = &
\alpha_V(Q^{*})
+ 2{\alpha_V^2(Q^{*}) \over \pi}
+ 15.728{\alpha_V^3 \over \pi^2} .
\end{eqnarray}
The one-loop coefficient is the same as in our previous paper~\cite{bgmr}, but
the two-loop one is changed due to the new result by 
Schr\"oder~\cite{Schroder}.  
However, the problem brought up in our
previous paper regarding the anomalous contribution with a color factor
proportional to $C_FN_C$ is still there. 
This type of color factor corresponds to
an iteration of the potential and thus cannot be part of the potential
itself. The origin of this contribution is not clear,
but it is probably an effect of the breaking of conformal invariance by the
$\overline{\mbox{MS}}$ scheme.
It should also be remarked that the conformal two-loop coefficient
between the  $\overline{\mbox{MS}}$ scheme and the $\alpha_V$ scheme
is large, indicating that there are large corrections between the two schemes.
This is of great importance for observables like heavy quark production close
to threshold where the next-to-next-to leading order correction is known
to be large in the $\overline{\mbox{MS}}$ scheme \cite{hoang_teubner}.
As another example the conformal coefficients for $\alpha_{\Gamma}^{NS}(s)$
in terms of $\alpha_{\overline{\mbox{\tiny MS}}}$ are also large,
\begin{eqnarray}  
\alpha_{\Gamma}^{NS}(s) & = &
\alpha_{\overline{\mbox{\tiny MS}}}^{(N_L)}(\mu^*)
+0.0833
\frac{\left(\alpha_{\overline{\mbox{\tiny MS}}}^{(N_L)}(\mu^*)\right)^2}{\pi} 
-23.2
\frac{\left(\alpha_{\overline{\mbox{\tiny MS}}}^{(N_L)}(\mu^*)\right)^3}{\pi^2}
\end{eqnarray}
where $\mu^*$ is the commensurate scale between $\alpha_{\Gamma}^{NS} $
and the $\overline{\mbox{MS}}$ scheme.
This should be compared with the  conformal relation between
$\alpha_{\Gamma}^{NS}$ and $\alpha_V$ where 
the coefficients are not as large as indicated in Eq.~(\ref{eq:agvfull}).
Thus it is better to relate observables directly without using
the intermediate analytic extension of the $\overline{\mbox{MS}}$ scheme.

The commensurate scale $Q^*$ between the $\overline{\mbox{MS}}$ and V
scheme is to next-to-leading order given by,
\begin{eqnarray}\label{eq:sing}
Q^* & = & Q\exp\left[\frac{5}{6}+
\frac{\displaystyle 
  \left[\left(\frac{35}{32}-\frac{3}{2}\zeta_3\right)C_F
  -\left(\frac{19}{48}-\frac{7}{4}\zeta_3\right)N_C
  \right] \psi_V^{(0)}(Q^*)\frac{\alpha_V(Q^*)}{\pi}  }
{\displaystyle\psi_V^{(0)}(Q^*) +  \psi_V^{(1)}(Q^*)\frac{\alpha_V(Q^*)}{\pi}}
\right]
\nonumber \\
& = &  Q\exp\left[0.833+
\frac{[7.314-0.443N_F^{(0)}(Q^*)] \alpha_V(Q^*) }
{5.500-0.333N_F^{(0)}(Q^*)+[4.058 -0.504N_F^{(1)}(Q^*)] \alpha_V(Q^*) }\right] .
\end{eqnarray}
In the limit $\alpha_V \to \infty$
the  scale $Q^*$ becomes large (for $N_F=3$, $Q^*=24Q$ in the limit
$\alpha_V \to \infty$). If the coupling freezes or is bounded 
for small scales, then this limit is of course not important.

\subsection{Definition of the Analytic
\mbox{${\widetilde{\alpha}_{\overline{\mbox{\tiny MS}}}}$} }

The definition of the analytic
\mbox{${\widetilde{\alpha}_{\overline{\mbox{\tiny MS}}}}$}
is based on generalizing Eq.~(\ref{eq:csrmsofv}) to be valid for all $Q$ by
using the mass-dependent $\alpha_V(Q,m_i)$,
\begin{eqnarray}\label{eq:amsdef}
\widetilde{\alpha}_{\overline{\mbox{\tiny MS}}}(Q,m_i)
 & \equiv &
\alpha_V(Q^{*},m_i) +  \frac{2}{3}N_C{\alpha_V^2(Q^{*},m_i) \over \pi},
\end{eqnarray}
with $m_i$ being the pole-masses for the quarks which do not depend on $Q$.
In addition we use the mass-dependent coupling $\alpha_V(Q,m_i)$
and the mass-dependent coefficients of the $\Psi_V$-function,
$\psi_V^{(0)}(Q,m_i)$ and $\psi_V^{(1)}(Q,m_i)$, in the
formula for $Q^{*}$ given by Eq.~(\ref{eq:sing}).
In the above definition we have only included terms to the order
which we are working, i.e. next-to-leading order, since
the effects from higher order terms on $Q^*$ is unknown. When going
to even higher orders, the relation between the analytic
\mbox{${\widetilde{\alpha}_{\overline{\mbox{\tiny MS}}}}$}
and the $\alpha_V$ scheme will contain large corrections as indicated in
Eq.~(\ref{eq:csrmsofv}), reflecting the
underlying large difference between the $\overline{\mbox{MS}}$ and $\alpha_V$
schemes.

We can also derive the $\Psi$-function for
\mbox{${\widetilde{\alpha}_{\overline{\mbox{\tiny MS}}}}$}
by taking the logarithmic derivative of Eq.~(\ref{eq:amsdef})
with respect to $Q$. This gives
\begin{equation}\label{eq:psiamsdef}
\widetilde {\Psi}_{\overline{\mbox{\tiny MS}}}(Q,m_i)
\equiv \Psi_V(Q^*,m_i) + 2 \frac{2N_C}{3} {\alpha_V(Q^{*},m_i)\over\pi}
\Psi_V(Q^{*},m_i),
\end{equation}
Re-expanding both sides of Eq.~(\ref{eq:psiamsdef}) in $\alpha_V$ using
Eq.~(\ref{eq:amsdef}) and equating order by order gives the first
two terms in $\widetilde {\Psi}_{\overline{\mbox{\tiny MS}}}(Q,m_i)$,
\begin{eqnarray}\label{eq:psi0def}
\widetilde {\psi}_{\overline{\mbox{\tiny MS}}}^{(0)}(Q,m_i)
& = & \psi_V^{(0)}(Q^*,m_i)  \\ \label{eq:psi1def}
\widetilde {\psi}_{\overline{\mbox{\tiny MS}}}^{(1)}(Q,m_i)
& = & \psi_V^{(1)}(Q^*,m_i)
\end{eqnarray}
reflecting the non-trivial mass dependence of the
${\Psi}$-function. We note
that when finite quark masses are included the first two terms in the
${\Psi}$-function are no longer universal but scheme dependent.

\begin{figure}[htbp]
\begin{center}
\mbox{\epsfig{figure=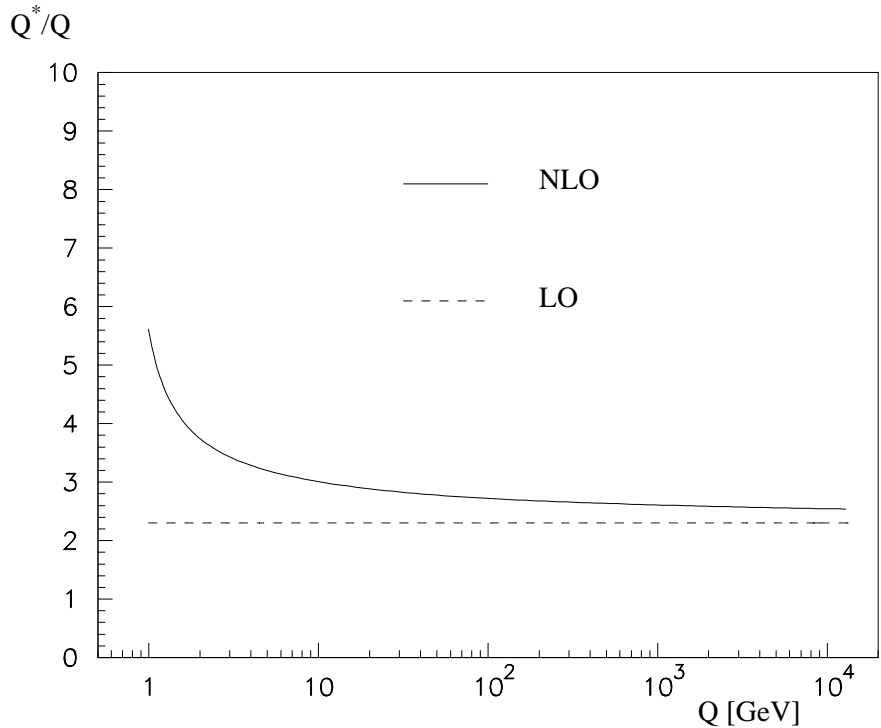,width=12cm}}
\end{center}
\caption[*]{The ratio of the commensurate scale $Q^*$ to $Q$
for the analytic extension of the $\overline{\mbox{MS}}$ scheme
as a function of $Q$ to leading (dashed) and next-to-leading (solid)
order.}
\label{fig:qratams}
\end{figure}

We thus arrive at the following evolution equation for
the analytic coupling $\widetilde{\alpha}_{\overline{\mbox{\tiny MS}}}(Q,m_i)$,
\begin{equation}\label{eq:amsrge}
{d \, \widetilde{\alpha}_{\overline{\mbox{\tiny MS}}}(Q,m_i) \over d \ln Q} =
- \widetilde{\psi}_{\overline{\mbox{\tiny MS}}}^{(0)}(Q,m_i)
 {\widetilde{\alpha}_{\overline{\mbox{\tiny MS}}}^2(Q,m_i) \over \pi}
- \widetilde{\psi}_{\overline{\mbox{\tiny MS}}}^{(1)}(Q,m_i)
 {\widetilde{\alpha}_{\overline{\mbox{\tiny MS}}}^3(Q,m_i) \over \pi^2} ,
\end{equation}
where $\widetilde{\psi}_{\overline{\mbox{\tiny MS}}}^{(0)}(Q,m_i)$
and  $\widetilde{\psi}_{\overline{\mbox{\tiny MS}}}^{(1)}(Q,m_i)$ are
given by Eqs.~(\ref{eq:psi0def}) and (\ref{eq:psi1def}), respectively,
and $m_i$ are the pole masses of the quarks.
One complication which arises when solving the evolution equation
is that the scale $Q^*$ has to be obtained recursively
since Eq.~(\ref{eq:sing}) contains $Q^*$ also on the right hand side.
In addition the approximation
$\alpha_V(Q^*,m_i)=\widetilde{\alpha}_{\overline{\mbox{\tiny MS}}}(Q,m_i)$
was used in the right hand side of Eq.~(\ref{eq:sing})
when solving the evolution equation for 
$\widetilde{\alpha}_{\overline{\mbox{\tiny MS}}}(Q,m_i)$.
The evolution equation was solved for numerically using the classical
Runge-Kutta algorithm.  The resulting scale $Q^*$
 calculated using Eq.~(\ref{eq:sing}) is shown in Fig.~\ref{fig:qratams}.

\begin{figure}[htbp]
\begin{center}
\mbox{\epsfig{figure=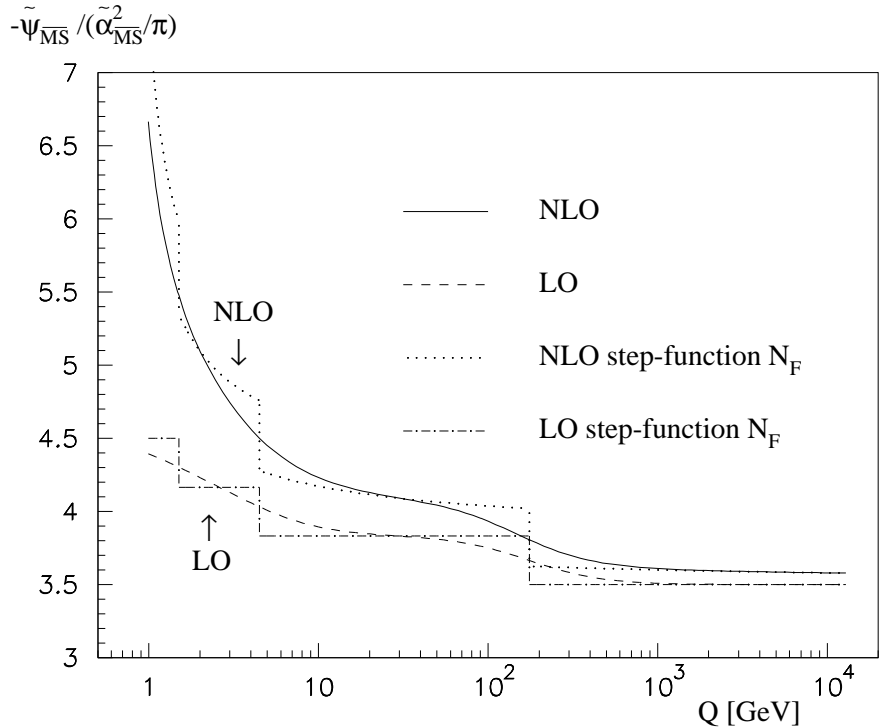,width=12cm}}
\end{center}
\caption[*]{The normalized $\Psi$-function,
$-\widetilde {\Psi}_{\overline{\mbox{\tiny MS}}}(Q,m_i)/
(\widetilde{\alpha}^2_{\overline{\mbox{\tiny MS}}}(Q,m_i)/\pi)$
in the analytic extension of the $\overline{\mbox{MS}}$ scheme
compared to the  conventional $\beta$-function with
discrete theta-function thresholds. The leading order results are shown
as a dashed and dot-dashed curve respectively whereas the
next-to-leading order results  are shown
as a solid curve and   dotted curve respectively.}
\label{fig:psiams}
\end{figure}

With the solution of the renormalization group equation for
$\widetilde{\alpha}_{\overline{\mbox{\tiny MS}}}$ we also
obtain the $\Psi$ function for the analytic extension of
$\overline{\mbox{MS}}$ which is shown in Fig.~\ref{fig:psiams}.
From the figure we also see that the $\Psi$-function in the
analytic approach and in the massless step-function approach,
with matching at the quark masses, follow each other closely
except for small scales where they start to deviate.

\subsection{Comparing the Analytic
$\widetilde{\alpha}_{\overline{\mbox{\tiny MS}}}$
with $\alpha_{\overline{\mbox{\tiny MS}}}$}

\begin{figure}[htb]
\begin{center}
\mbox{\epsfig{figure=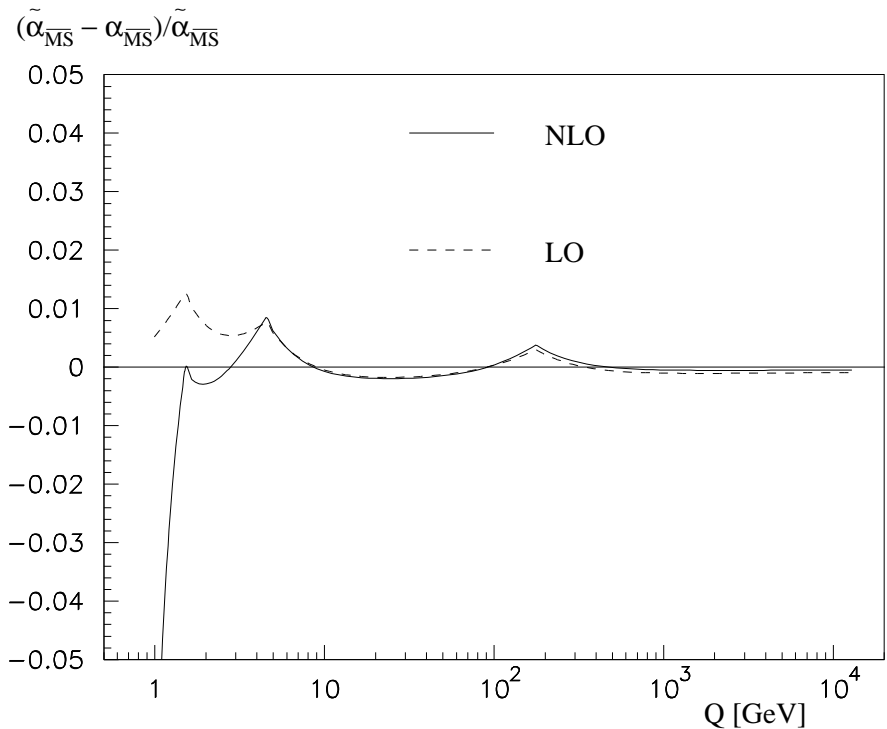,width=12cm}}
\end{center}
\caption[*]{The relative difference between the
solutions to
the 2-loop renormalization group equation using the analytic
$\Psi$-function,
$\widetilde{\alpha}_{\overline{\mbox{\tiny MS}}}(Q)$, and conventional
discrete theta-function thresholds,
$\alpha_{\overline{\mbox{\tiny MS}}}(Q)$, with matching at quark masses.
The solid curve shows the next-to-leading order  result.
For comparison the 1-loop result is shown as a dashed curve.
The solutions have been
obtained numerically starting from the world average \protect\cite{Burrows}
$\alpha_{\overline{\mbox{\tiny MS}}}(M_Z) = 0.118$.}
\label{fig:adiffnlo}
\end{figure}

We now compare the analytic
$\widetilde{\alpha}_{\overline{\mbox{\tiny MS}}}$
with the conventional discrete theta-function treatment of flavor
thresholds with matching at quark masses,
$\alpha_{\overline{\mbox{\tiny MS}}}$. The relative difference between the
two is shown in Fig.~\ref{fig:adiffnlo} both in leading and next-to leading
order.

As can be seen from the figure, the difference between the
analytic and conventional treatment of quark masses in the running persists
when going from leading to next-to-leading
order. In fact we expect this difference to remain
to all orders. The reason is that the $\Psi$ function is
not continuous in the massless approach and the stepsize at the thresholds
is governed by the lowest order term
$\psi^{(0)}$. Thus
there will always be a finite difference between the continuous
$\Psi$-function and the one with theta-function thresholds. It is
important to recognize that this feature is not eliminated by the
fact that when going
to higher orders the dependence on the matching scale in the massless
approach becomes smaller. The only way to include these mass effects in
the ordinary $\overline{\mbox{MS}}$ treatment is by making a higher twist
analysis to all orders in $m^2/Q^2$ and $Q^2/m^2$.



\begin{thebibliography}{999}

\bibitem{Susskind}
L.~Susskind, {\it Coarse grained quantum chromodynamics},
lectures given at Les Houches 1976,
in R.~Balin and C.~H.~Llewellyn Smith (eds.),
{\it Weak and Electromagnetic Interactions At High Energies},
(North-Holland 1977, 207-308).

\bibitem{derujula}
A.~De R{\'u}jula and H.~Georgi, Phys.~Rev.~{\bf D13}, 1296 (1976).

\bibitem{Georgi_Politzer}
H.~Georgi and H.D.~Politzer, Phys.~Rev.~{\bf D14}, 1829 (1976).

\bibitem{Ross}
D.A.~Ross, Nucl.~Phys.~{\bf B140}, 1 (1978);
T.~Goldman and D.A.~Ross, Nucl.~Phys.~{\bf B171}, 273 (1980).

\bibitem{shirkov}
D.V.~Shirkov,
Theor. Math. Phys. {\bf 93}, 1403 (1992).
See also K.A.~Milton, O.~P. ~Solovtsova,
Phys. Rev. {\bf D57}, 5402 (1998), and references therein.


\bibitem{chyla}
J.~Ch{\'y}la, Phys.~Lett.~{\bf B 351}, 325 (1995).

\bibitem{yh} T.~Yoshino, K.~Hagiwara,
Z.Phys. {\bf C 24}, 185 (1984).

\bibitem{jt} F.~Jegerlehner, O.V.~Tarasov,
hep-ph/9809485 and DESY 98-093.

\bibitem{ac} T.~Appelquist, J.~Carazzone,
Phys.~Rev.~{\bf D11}, 2856 (1975).

\bibitem{csr}
S.J.~Brodsky, H.J.~Lu, Phys.~Rev.~{\bf D51}, 3652 (1995);
hep-ph/9506322.

\bibitem{melles98} 
M.~Melles, Phys.~Rev.~{\bf D58}:114004, 1998.

\bibitem{vegas} 
G.P.~Lepage, J.~Comp.~Phys.~{\bf 27}, 192 (1978);
Cornell preprint, CLNS-80/447, March 1980.

\bibitem{bgmr}
S.J.~Brodsky, M.S.~Gill, M.~Melles, J.~Rathsman,
Phys.~Rev.~{\bf D58}:116006, 1998.

\bibitem{blm}
S.J.~Brodsky, G.P.~Lepage and P.B.~Mackenzie,
Phys.~Rev.~{\bf D28}, 228 (1983).

\bibitem{Frishman}
S. J.~Brodsky, Y.~Frishman and G. P.~Lepage,
Phys. Lett. {\bf 167B}, 347 (1986);
S. J.~Brodsky, P.~Damgaard, Y.~Frishman and G. P.~Lepage,
Phys. Rev. {\bf D33}, 1881 (1986).


\bibitem{Muller}
D.~Muller, Phys. Rev. {\bf D59}, 116003 (1999); A. V.~Belitsky and D.~Muller,
Nucl. Phys. {\bf B537}, 397 (1999); D.~Muller, Phys. Rev. {\bf D49}, 2525
(1994).

\bibitem{Blumlein}
J.~Bl{\"u}mlein, V.~Ravindran and W.L.~van Neerven, 
Acta Phys.~Polon.~{\bf B29}, 2581 (1998).

\bibitem{Fischler}
W.~Fischler, Nucl.~Phys.~{\bf B129}, 157 (1977).

\bibitem{Appelquist_Dine_Muzinich}
T.~Appelquist, M.~Dine, I.J.~Muzinich,
Phys.~Lett.~{\bf 69B}, 231 (1977); Phys.~Rev.~{\bf D17}, 2074 (1978).

\bibitem{Feinberg}
F.L.~Feinberg, Phys.~Rev.~Lett~{\bf 39}, 316 (1977);
Phys.~Rev.~{\bf D17}, 2659 (1978);
S.~Davis, F.L.~Feinberg, Phys.~Lett.~{\bf 78B}, 90 (1978).

\bibitem{Billoire}
A.~Billoire, Phys.~Lett.~{\bf 92B}, 343 (1980).

\bibitem{Peter}
M.~Peter, Phys.~Rev.~Lett.~{\bf 78}, 602 (1997);
Nucl.~Phys.~{\bf B 501}, 471 (1997).

\bibitem{Schroder}
Y.~Schr\"oder,  Phys.~Lett.~{\bf B447}, 321 (1999).

\bibitem{mspole}
R.~Tarrach, Nucl.~Phys.~{\bf B183}, 384 (1981).

\bibitem{gl} M.~Gell-Mann, F.E.~Low,
Phys. Rev. {\bf 95}, 1300 (1954).

\bibitem{ks} G.~K\"allen, A.~Sabry,
Dan.~Mat.~Fys.~Medd.~{\bf 29}, No.17 (1955).

\bibitem{br} R.~Barbieri, E.~Remiddi,
Nuovo Cimento {\bf 13 A}, 99 (1973).

\bibitem{k} B.A.~Kniehl,
Nucl. Phys. {\bf B 347}, 65 (1990).

\bibitem{Hoang}
A.H.~Hoang, M.~Je\.zabek, J.H.~K\"uhn, T.~Teubner,
Phys.~Lett.~{\bf B 338}, 330 (1994).

\bibitem{Chetyrkin}
K.G.~Chetyrkin, Phys.~Lett.~{\bf B 307}, 169 (1993).

\bibitem{Soper_Surguladze}
D.E.~Soper and L.R.~Surguladze, Phys.~Rev.~Lett.~{\bf 73}, 2958 (1994).

\bibitem{L_R_V_NPB}
S.A.~Larin, T.~van Ritbergen, J.A.M.~Vermaseren,
Nucl.~Phys.~{\bf B438}, 278 (1995).

\bibitem{grunberg}
G. Grunberg,  Phys.~Lett.~{\bf B95}, 70 (1980),
 Phys.~Lett.~{\bf B11}, 501 (1982),
Phys.~Rev.~{\bf D29}, 2315 (1984).

\bibitem{msmatch}
S.~Weinberg, Phys.~Lett. {\bf 91B}, 51 (1980);
L.~Hall, Nucl.~Phys.~{\bf B178}, 75 (1981);
P.~Binetruy, T.~Schucker, Nucl.~Phys~{\bf B178}, 293, 307 (1981).

\bibitem{Kataev}
S.J.~Brodsky, G.T.~Gabadadze, A.L.~Kataev, H.J.~Lu,
Phys.~Lett.~{\bf B372}, 133 (1996).

\bibitem{Mueller}
A.H.~Mueller, Phys.~Lett.~{\bf B308}, 355 (1993).


\bibitem{hoang_teubner}
A.H.~Hoang and T.~Teubner, Phys.~Rev.~{\bf D58}, 114023 (1998).

\bibitem{Burrows}
P.N.~Burrows, Acta Phys.~Polon.~{\bf B28}, 701 (1997).

\end{thebibliography}
\end{document}